\documentclass[aps,pra,twocolumn,10pt,superscriptaddress,longbibliography, floatfix,amsmath,amssymb]{revtex4-1}
\usepackage{graphicx}
\usepackage{xcolor}
\usepackage{ulem}

\newcommand\m[1]

\begin{document}


\title{Generation of intraparticle quantum correlations in amplitude damping channel and its robustness}

\author{Animesh Sinha Roy}
\affiliation{Raman Research Institute, C. V. Raman Avenue, Sadashivanagar, Bengaluru, Karnataka 560080, India}

\author{Namitha C.V. }
\affiliation{Raman Research Institute, C. V. Raman Avenue, Sadashivanagar, Bengaluru, Karnataka 560080, India}

\author{Subroto Mukerjee}
\affiliation{Indian Institute of Science,  C. V. Raman Road, Bengaluru, Karnataka 560012, India}

\author{Prasanta K. Panigrahi}
\affiliation{Department of Physical Sciences, Indian Institute of Science Education and Research Kolkata, Mohanpur, Nadia, 741246, West Bengal, India}

\author{Urbasi Sinha }
\email[]{usinha@rri.res.in}
\affiliation{Raman Research Institute, C. V. Raman Avenue, Sadashivanagar, Bengaluru, Karnataka 560080, India}






\date{\today}

\begin{abstract}

Quantum correlations between two or more different degrees of freedom of the same particle are sometimes referred to as intraparticle entanglement. In this work, we study these intra-particle correlations between two different degrees of freedom under various decoherence channels, viz. amplitude damping, depolarizing and phase damping channels. We mainly focus on the amplitude damping channel for which we obtain an exact analytical expression for the concurrence of an arbitrary initial pure state. In this channel, we observe the unique feature of entanglement arising from a separable initial state. We also show that this channel allows for a revival of entanglement with increasing damping parameter including from a zero value of the concurrence. We also consider the amplitude damping channel for interparticle entanglement and show that it does not display any of the above-mentioned interesting features. Further, for comparable parameters, the decay of entanglement in the interparticle system is much greater than in the intraparticle system, which we also find to be true for the phase damping and depolarizing channels. Thus, intraparticle entanglement subjected to damping is much more robust than interparticle entanglement.
\end{abstract}

\maketitle

    
\section{\label{sec:level1}Introduction}
Entanglement can arise between two spatially separated particles or two different degrees of freedom of the same particle due to the nonseparability of the states in the Hilbert space. 
Interparticle entanglement refers to quantum correlations between two or more spatially separated particles in a single degree of freedom \cite{Horodecki09}, while intraparticle entanglement refers to quantum correlations between two or more degrees of freedom of the same particle \cite{Azzini20, Basu}. Intraparticle entanglement has been experimentally verified for various quantum systems \cite{Michler2000,Gadway2009,Barreiro2005,Hasegawa2003,Shen2020}. It is an important property for the testing of \textit{quantum contextuality} and other various applications, such as quantum key distribution protocols \cite{Sun2011,Adhikari2015}, quantum teleportation protocols \cite{Heo2015,Pramanik2010,Hong2015}, entanglement swapping protocols \cite{Adhikari2010} etc.
      
Any real-life applications of entanglement would suffer from interactions with a noisy environment. Thus, for realistic applications, it is imperative to investigate the nature of intraparticle entanglement in noisy channels. Further, to find its efficacy, the same should be contrasted and compared with the decohering effects of noisy channels on interparticle entanglement, which have been systematically studied \cite{Yu2009,Alm2007,Almeida2008}. Investigating intraparticle entanglement in noisy channels and comparing it with interparticle entanglement will help researchers in developing more robust quantum protocols and systems. This research is crucial for advancing applications like quantum key distribution, teleportation, and entanglement swapping, ensuring they can function effectively in real-world scenarios.

In this work, we study the effects of noisy channels, viz. amplitude damping, phase damping and depolarizing channels, on the intraparticle correlations shared between degrees of freedom of a single particle. We study the entanglement between two degrees of freedom by calculating the concurrence as a function of the channel and input parameters. We mainly focus on the amplitude damping channel, for which we obtain an exact analytical expression for the concurrence and exhibit the very interesting phenomenon of the rebirth of entanglement as a function of the channel parameter. This demonstrates that increasing noise, as quantified by the channel parameter, can counter-intuitively be beneficial for the generation of entanglement. This is the central result of our work and offers a clear distinction between the effect of damping on intraparticle entanglement and the more conventional case of interparticle entanglement with non-Markovian noise \cite{BBellomo}. We explain the behavior of intraparticle entanglement through an extensive analysis of the concurrence in terms of the channel and input state parameters. Finally, we investigate the decoherence of intraparticle entanglement in the phase damping and depolarizing channels and contrast it with the interparticle entanglement scenario to demonstrate the robustness of intraparticle entanglement. 

We model the damping using the Kraus operator formalism \cite{Kraus83}, where the evolution of an initial state $\rho$ is described by a trace preserving map $\rho \rightarrow \rho^{'}=\sum_{i} M_{i}\rho M_{i}^{\dagger}$. ${M}_{i}$ are the Kraus operators satisfying the completeness relation $\sum_{i} M_{i}^{\dagger} M_{i}=\mathcal{I}$. 
      
We consider a pure state of single quantum system $S$ with two different degrees of freedom, $A$ and $B$ whose states belong to two dimensional Hilbert spaces $\mathcal{H}_{A}$ and $\mathcal{H}_{B}$ with orthonormal bases $\{|a_{1}\rangle$, $|a_{2}\rangle \}$ and $\{|b_{1}\rangle$, $|b_{2}\rangle \}$ respectively:
 \begin{eqnarray}
 \label{eqn:inputstateMain}
|\psi\rangle_{in}=a|a_{1}b_{1}\rangle +b|a_{1}b_{2}\rangle +c|a_{2}b_{1}\rangle +d|a_{2}b_{2}\rangle
\end{eqnarray}
with complex coefficients $a$, $b$, $c$ and $d$ with $|a|^{2}+|b|^{2}+|c|^{2}+|d|^{2}=1$. We note that this is identical to the form of a pure state of two spatially separated particles, each with states in separate two dimensional Hilbert spaces. We can thus use the formalism developed to calculate the concurrence of such a system for our system as well \cite{Wootters1998}. For an arbitrary bipartite quantum state $\rho$, the concurrence is defined as
\begin{eqnarray}
 \label{eqn:concurrence1}
C(\rho) = max\{ 0, \sqrt{\lambda_{1}}- \sqrt{\lambda_{2}}- \sqrt{\lambda_{3}}- \sqrt{\lambda_{4}}\}
\end{eqnarray}
where the quantities $\lambda_{i}$ are the eigenvalues in decreasing order of the matrix:
\begin{eqnarray}
 \label{eqn:concurrence2}
R = \rho (\sigma_{y}\otimes\sigma_{y})\rho^{*}(\sigma_{y}\otimes\sigma_{y}).
\end{eqnarray}
 Here $\rho^{*}$ is the complex conjugate of the matrix representing the input state $\rho$ and $\sigma_{y}$ is the Pauli matrix,
 \begin{eqnarray*}
\label{eqn:concurrence3}
 \sigma_{y} = \begin{pmatrix}
0 & -i \\
i & 0
\end{pmatrix}
\end{eqnarray*}
The intraparticle entanglement between two degrees of freedom for the pure state in Eqn.~\ref{eqn:inputstateMain} is $C = 2|ad-bc|$ . 

The evolution of the concurrence may show significant differences depending on the decoherence of the respective channels. For an interparticle entangled state, the environment acts locally on each particle \cite{KarlKraus, Almeida2008}. The Kraus operators for the system are thus tensor products of the Kraus operators for each particle, each dimension equal to that of the local Hilbert space of the particle. For an intraparticle entangled state, both degrees of freedom belong to the same particle and it is natural to consider an environment that acts on both degrees of freedom simultaneously. The Kraus operators are, in general, not separable into operators for each degree of freedom, and their dimension equals that of the Full Hilbert space of the particle.

In the case of intraparticle entanglement, since two degrees of freedom belong to the same particle, we consider a noise model where noise looks at the particle as a single entity and acts on the state of the particle as the quantum state of that single entity. Since the state of this single entity belongs to a higher four-dimensional system, i.e., a qudit (d=4), the noise
also acts on the qudit. In this model, noise cannot detect each degree of freedom individually. That is why we consider the noise model where the Kraus operators belong to a higher dimensional system. If on the other hand, a noise model can detect each degree of freedom separately, then the dimension of that noise acting on each degree of freedom is the same as the dimension of the state of each degree of freedom. In that case, the total noise acting on the system is nothing but the tensor product of the noise acting on each degree of freedom. In such a situation, the effect of noise on an intraparticle entangled state and an interparticle entangled state would turn out to be the same. In our work, we do not consider such noise models and stick to the more physically motivated scenario exemplified by the noise acting on the particle as a whole.

\section{\label{sec:level5}Amplitude damping channel}
The amplitude damping channel represents a physical process that dissipates energy to the environment of a system via interactions. The effect of amplitude damping noise on an intraparticle entangled state $|\psi\rangle_{in}$, defined by Eqn.~\ref{eqn:inputstateMain}, can be represented by the following the Kraus operators \cite{ArijitDutta, AlejandroFonseca}
\begin{eqnarray*}
M_{0} &=&  |0\rangle\langle 0|+\sqrt{1-P}\sum_{j=1}^{3}|j\rangle\langle j|~,\\
M_{i} &=& \sqrt{P} |0\rangle\langle i| ~~(i=1,2,3)
\end{eqnarray*}
where $P$ is the channel parameter and the states $|0 \rangle=|a_1b_1\rangle$, $|1 \rangle=|a_1b_2\rangle$, $|2 \rangle=|a_2b_1\rangle$ and $|3 \rangle=|a_2b_2\rangle$ for brevity. The state $|0\rangle$ corresponds to the ground state of the system, and since the damping affects both degrees of freedom simultaneously, we assume that 
it causes transitions from all excited states to the ground state in an equivalent manner. After the evolution of $|\psi\rangle_{in}$ through the amplitude damping channel, the final state becomes (Appendix \ref{sec:appendix1})
\begin{eqnarray}
 \label{eqn:outputstateintra}
&\rho_{out} = \big[P+(1-P)|a|^{2}\big]|a_{1}b_{1}\rangle\langle a_{1}b_{1}| + \sqrt{1-P}\sqrt{1-|a|^{2}}\nonumber\\
&\times\big(a|a_{1}b_{1}\rangle\langle \phi |+a^{*}|\phi\rangle\langle a_{1}b_{1}|\big) +(1-P)(1-|a|^{2})|\phi\rangle\langle\phi |
\end{eqnarray}
where $a^{*}$ is the complex conjugate of the state parameter $a$ and
\begin{eqnarray}
|\phi\rangle = \frac{1}{\sqrt{1-|a|^{2}}} \Big(b|a_{1}b_{2}\rangle +c|a_{2}b_{1}\rangle +d|a_{2}b_{2}\rangle \Big)
\end{eqnarray}
Since $\rho_{out}$ can be written in terms of two orthogonal states $|a_{1}b_{1}\rangle$ and $|\phi\rangle$, it has two nonzero eigenvalues. As a result the matrix $R$ [Eq.~(\ref{eqn:concurrence2})] also has two nonzero eigenvalues, which are
\begin{eqnarray}
\lambda_{\pm} = \frac{1}{2}(S \pm \sqrt{S^{2}-T^{2}})
\end{eqnarray}
where (Appendix \ref{sec:appendix1})
\begin{eqnarray}
S &=& 4|b|^{2}|c|^{2}(1-P)^{2}-4(1-P)^{3/2}[adb^{*}c^{*}+bca^{*}d^{*}]\nonumber\\
& & + 2|d|^{2}(1-P)[P+(1-P)|a|^{2}],\nonumber\\
T &=& 2P(1-P)(1-|a|^{2})|d|^{2}
\end{eqnarray}
After substituting the expressions for $S$ and $T$ into $\lambda_{\pm}$ and some algebra, we get  the concurrence of the output state as (Appendix \ref{sec:appendix1})
\begin{eqnarray}
    C = 2\big|bc\sqrt{(1-P)}-ad\big|\sqrt{1-P}
    \label{eqn:concurrence_pure}
\end{eqnarray}
The above expression is one of the key results of this paper. If there is no noise (i.e. $P=0$) $C =2|bc-ad|$, as expected for the initial pure state. The concurrence for $P \neq 0$ is equivalent to the substitution $\{a,b,c,d\} \rightarrow \{a,b\sqrt{1-P},c\sqrt{1-P},d\sqrt{1-P}\}$, reflecting the fact that amplitude damping causes the particle to remain in the excited states $|1\rangle$, $|2\rangle$ and $|3\rangle$ with probability $1-P$.

\begin{figure}[h]
\centering
\includegraphics[scale=0.8]{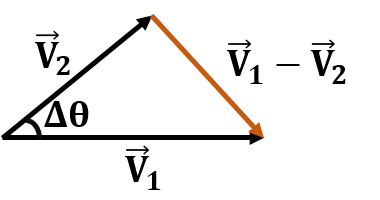}
\caption{In this figure $\vec{V}_{1}$, $\vec{V}_{2}$, $\vec{V}=\vec{V}_{1} - \vec{V}_{2}$ and $\Delta\theta $ are shown.}
\label{fig:vectorv}
\end{figure}

To better understand the behavior of the concurrence, we write
 \begin{eqnarray}
C &=& 2\Big[ |b|^{2}|c|^{2}(1-P)+|a|^{2}|d|^{2}\nonumber\\
&-& 2|a||b||c||d|\sqrt{1-P}\cos (\Delta\theta) \Big]^{1/2} \sqrt{1-P}
\end{eqnarray}
in terms of the amplitudes and phases of the coefficients $a$, $b$, $c$ and $d$, where $\Delta\theta = \theta_{b}+\theta_{c}-\theta_{a}-\theta_{d}$.
The concurrence can thus also be geometrically interpreted as $2\sqrt{1-P}$ times the length of the difference between two vectors $\vec{V}_1$ and $\vec{V}_2$ of lengths $|b||c|\sqrt{1-P}$ and $|a||d|$ respectively with angle $\Delta \theta$ between them, as shown in Fig.~\ref{fig:vectorv}.
It can be seen that the concurrence is zero when $P=1$ or
 \begin{eqnarray}
 \label{eqn:concurrence_2Main}
 P = 1 - \Big(\frac{|a||d|}{|b||c|}\Big)^{2}~~\text{and}~~ \Delta\theta =0 , \pm 2\pi.
 \end{eqnarray}
The condition $\Delta\theta =0 , \pm 2\pi$ arises since the individual angles ($\theta_a$, $\theta_b$, $\theta_c$, $\theta_d$) $\in [0,2\pi]$. It is important that for fixed values of the coefficients $a$, $b$, $c$ and $d$, the concurrence can go to zero only at a particular value of $P$ as opposed to a range of values. This is a consequence of the fact that two of the eigenvalues of the final state density matrix $\rho_{out}$ are equal to zero, and hence so are two of the eigenvalues of the matrix $R$ defined in Eqn.~\ref{eqn:concurrence2} used to calculate the concurrence. A zero value of concurrence results when the remaining two non-zero eigenvalues of $R$ become degenerate. This corresponds to a level crossing, which can occur only at isolated points in parameter space, and thus, the concurrence cannot be zero over a range of values of $P$. This behavior is unique to the intraparticle amplitude damping channel and is in sharp contrast to that observed in the other channels, as shown later. For the other channels, all four eigenvalues of $R$ are, in general, non-zero and thus, the concurrence can remain zero over a range of values of $P$ without exhibiting a revival.
\begin{figure}[h]
\centering
\includegraphics[scale=0.48]{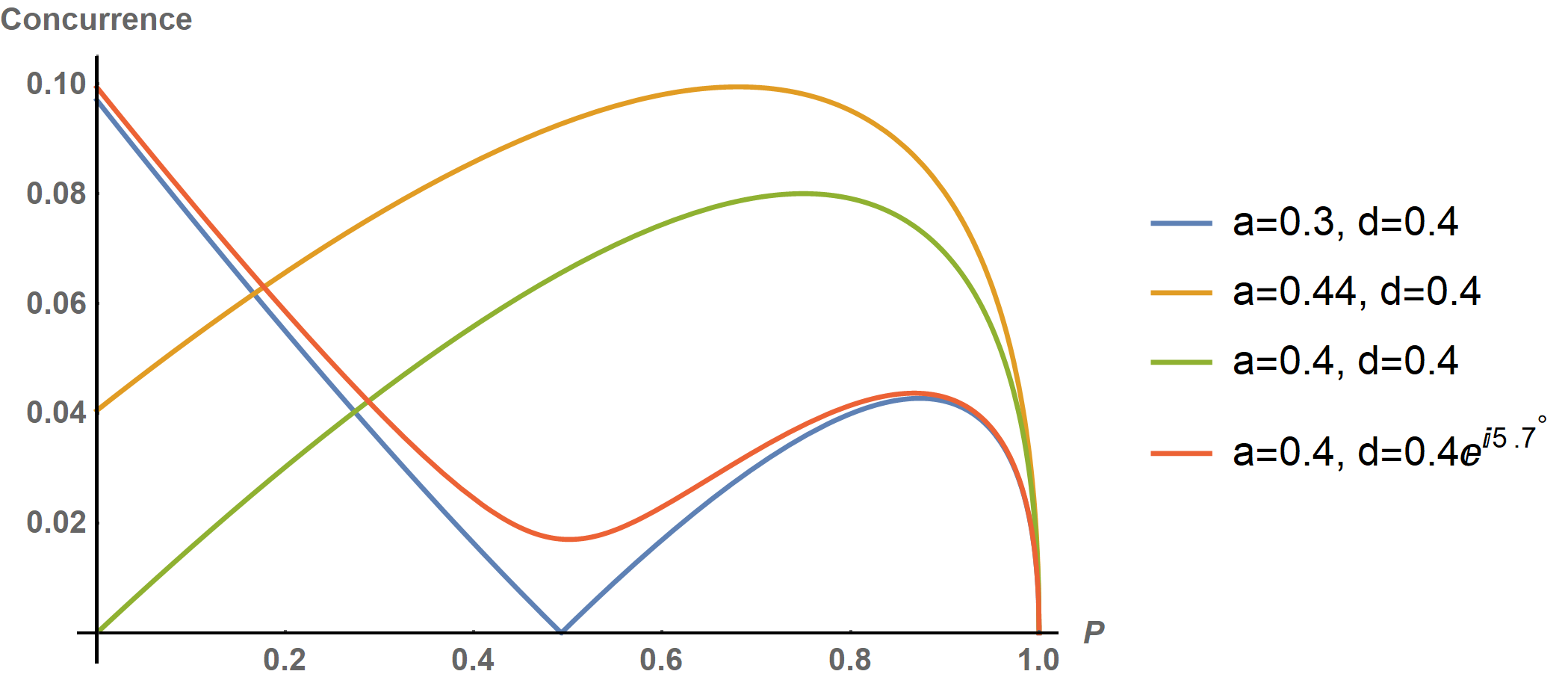}
\caption{The variation of the concurrence with $P$ for three different input states with state parameters $|a|=0.3$, $0.44$ and $0.4$ are represented by the blue, yellow and green plots. Other state parameters for these three input states are $|c|=0.2$,  $|d|=0.4$ and $\Delta\theta = 0$. The variation of the concurrence with channel parameter $P$ for an input state with state parameters $|a|=0.3$, $|b|=0.84$, $|c|=0.2$, $d=0.4$, $\theta_{a}=\theta_{b}=\theta_{c}=0$ and $\theta_{d}=5.7^{\circ}$ is represented by the red plot. The state representing the blue plot satisfies the condition (i) $|a||d|<|b||c|$, which first shows entanglement sudden death and then a rebirth of entanglement and finally asymptotically decays to zero. The value of $P$ where we observe the entanglement sudden death is given by Eqn.~\ref{eqn:concurrence_2Main}. The state representing the yellow plot satisfies the condition (ii) $|a||d|>|b||c|$, which first shows the creation of entanglement and finally asymptotically decays to zero. The state representing the green plot is initially a separable state satisfying the condition (iii) $|a||d|=|b||c|$ and with the increase of $P$, the concurrence first becomes nonzero and gradually increases and finally asymptotically decreases to zero. In the cases (i), (ii) and (iii) $\Delta\theta = 0$. If we make $\theta_{d} = 5.7^{\circ}$ while $\theta_{a}=\theta_{b}=\theta_{c}=0$, i.e., $\Delta\theta \neq 0 \pm 2\pi$, then we observe that the concurrence reaches its minimum nonzero value $C_{-}$ and then a maximum values $C_{+}$ and finally asymptotically decays to zero.}
\label{fig:intra1Main}
\end{figure}
The concurrence can go to zero only when the condition in Eqn.~\ref{eqn:concurrence_2Main} is satisfied. In Fig.~\ref{fig:intra1Main}, one can observe three distinct evolutions of the concurrence represented by the blue, yellow, and green plots. These evolutions correspond to three different input states with state parameters that satisfy Eqn.~\ref{eqn:concurrence_2Main}. For other combinations of parameters, it can still have a minimum at a value $P\neq 1$. This is also a type of revival, although not from the value of zero. The condition for such a minimum to exist is $\cos \Delta \theta \geq 0$ and $|\sin \Delta \theta| \leq 1/3$ \cite{Supplemental}. Since the concurrence has to go to zero at $P=1$ and is a smooth function of $P$, it has to go through a maximum as well. The values of $P$ at which the minima and maxima occur ($P_-$ and $P_+$ respectively) are (Appendix \ref{sec:appendix1}).
\begin{eqnarray}
\label{eqn:concurrence_8}
P_{\pm} = 1-\frac{|a|^{2}|d|^{2}}{16|b|^{2}|c|^{2}}\Big[3\cos\Delta\theta \mp \sqrt{(3\cos\Delta\theta)^{2}-8}\Big]^{2}
\end{eqnarray}
with $P_+ > P_-$ as expected. The corresponding minimum and maximum values ($C_-$ and $C_+$ respectively) of the concurrence are (Appendix \ref{sec:appendix1})
\begin{eqnarray}
\label{eqn:concurrence_9}
C_{\pm} &=& \frac{|a|^{2}|d|^{2}}{2\sqrt{3}|b||c|}\Big[ 1-\frac{\{3\cos\Delta\theta \mp \sqrt{(3\cos\Delta\theta)^{2}-8}~\}^{2}}{16}\Big]^{1/2}\nonumber\\
&\times &\Big[ 3\cos\Delta\theta \mp \sqrt{(3\cos\Delta\theta)^{2}-8} \Big]
\end{eqnarray}

In Fig.~\ref{fig:intra1Main}, the red plot illustrates the evolution of the concurrence for an intraparticle entangled state with state parameters that do not conform to equation (10), i.e., $\Delta\theta \neq 0,\pm 2\pi$. The non-monotonic behavior of the concurrence as a function of $P$ is special to the amplitude damping channel and can be quantified by $\tilde{C} = \frac{C_+-C_-}{C_++C_-}$. This can be taken to be a measure of the effectiveness of the channel in terms of reviving the concurrence. {$\tilde{C}$ is a maximum when $\Delta \theta=0, \pm 2\pi$. The behavior of $\Tilde{C}$ with $\Delta\theta$ is shown in Fig.~\ref{fig:ctildeMain}. The Fig.~\ref{fig:ctildeMain} shows that $\Tilde{C}$ is maximum at $\Delta\theta =0$ and decreases with the increase of $\Delta\theta$. $\Tilde{C}=0$ at $\Delta\theta = \cos^{-1}{\frac{2\sqrt{2}}{3}}= 19.47^{\circ}$.
\begin{figure}[h]
\centering
\includegraphics[scale=0.58]{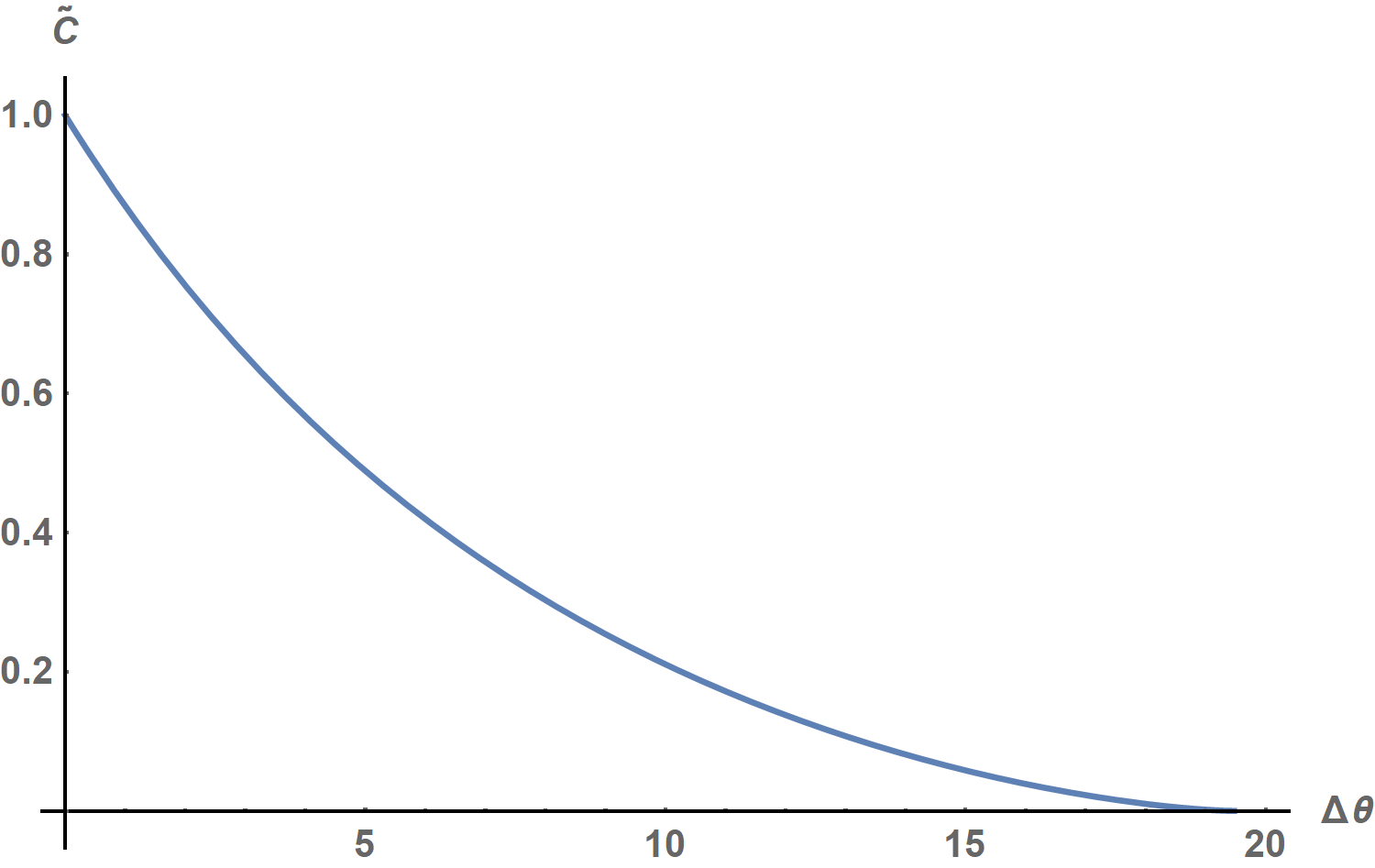}
\caption{Variation of $\Tilde{C}$ with $\Delta\theta$ under the effect of the amplitude damping channel. Here $\Tilde{C}$ is maximum at $\Delta\theta=0$. With the increase of $\Delta\theta$, $\Tilde{C}$ gradually decreases and becomes zero at $\Delta\theta = 19.47^{\circ}$.}
\label{fig:ctildeMain}
\end{figure}

 {\em Interparticle entanglement:} Next, for the sake of comparison, we consider the decoherence of an inter-particle entangled state $|\psi\rangle_{in} $ due to the amplitude damping channel. The two particles $A$ and $B$ are assumed to have individual two dimensional Hilbert spaces and are spatially well separated. The environment is assumed to act on each particle locally and identically. Each of the Kraus operators for the system is thus a tensor product of the local Kraus operators for each particle.
The local Kraus operators are \cite{Almeida2008}
\begin{eqnarray}
M_{0}=\begin{pmatrix}
  1 & 0\\
 0 & \sqrt{1-P}
\end{pmatrix}
~~~~,~~~~
      M_{1}=\begin{pmatrix}
  0 & \sqrt{P}\\
 0 & 0
\end{pmatrix}
      \end{eqnarray}

We consider a pure input state that has the same form as in Eqn.~\ref{eqn:inputstateMain}. The concurrence of the output state is
\begin{eqnarray}
\label{eqn:ampinterconcurrence}
C = max\Big[0,\sqrt{\alpha +\beta}-\sqrt{\alpha -\beta}-2d^{2}P(1-P)\Big]
\end{eqnarray} 
where
\begin{eqnarray}
\label{eqn:ampinterconcurrencealphabeta}
\alpha &=& 2(ad-bc)^{2}(1-P)^{2}+d^{4}P^{2}(1-P)^{2},\nonumber\\
\beta &=& 2(ad-bc)^{2}(1-P)^{2}\sqrt{1-\frac{d^{4}}{(ad-bc)^{2}}P^{2}}
\end{eqnarray}
where $P$ is the channel parameter. The effect of interparticle amplitude damping is quite distinct from that of intraparticle damping. Since no local operator can create entanglement \cite{Bennett1996, Vidal2000}, an initially unentangled state continues to remain unentangled under the effect of interparticle amplitude damping channel. Further, as shown in Fig.~\ref{fig:amplinter}, if the concurrence goes to zero at $P<1$, it does not exhibit a revival. Mathematically, this is related to all four eigenvalues of the matrix $R$, generically being non-zero, as remarked earlier. Further, the concurrence does not appear to exhibit a minimum and always decreases monotonically, showing that increasing the level of damping has a deleterious effect on the concurrence.  
\begin{figure}[h]
\centering
\includegraphics[scale=0.8]{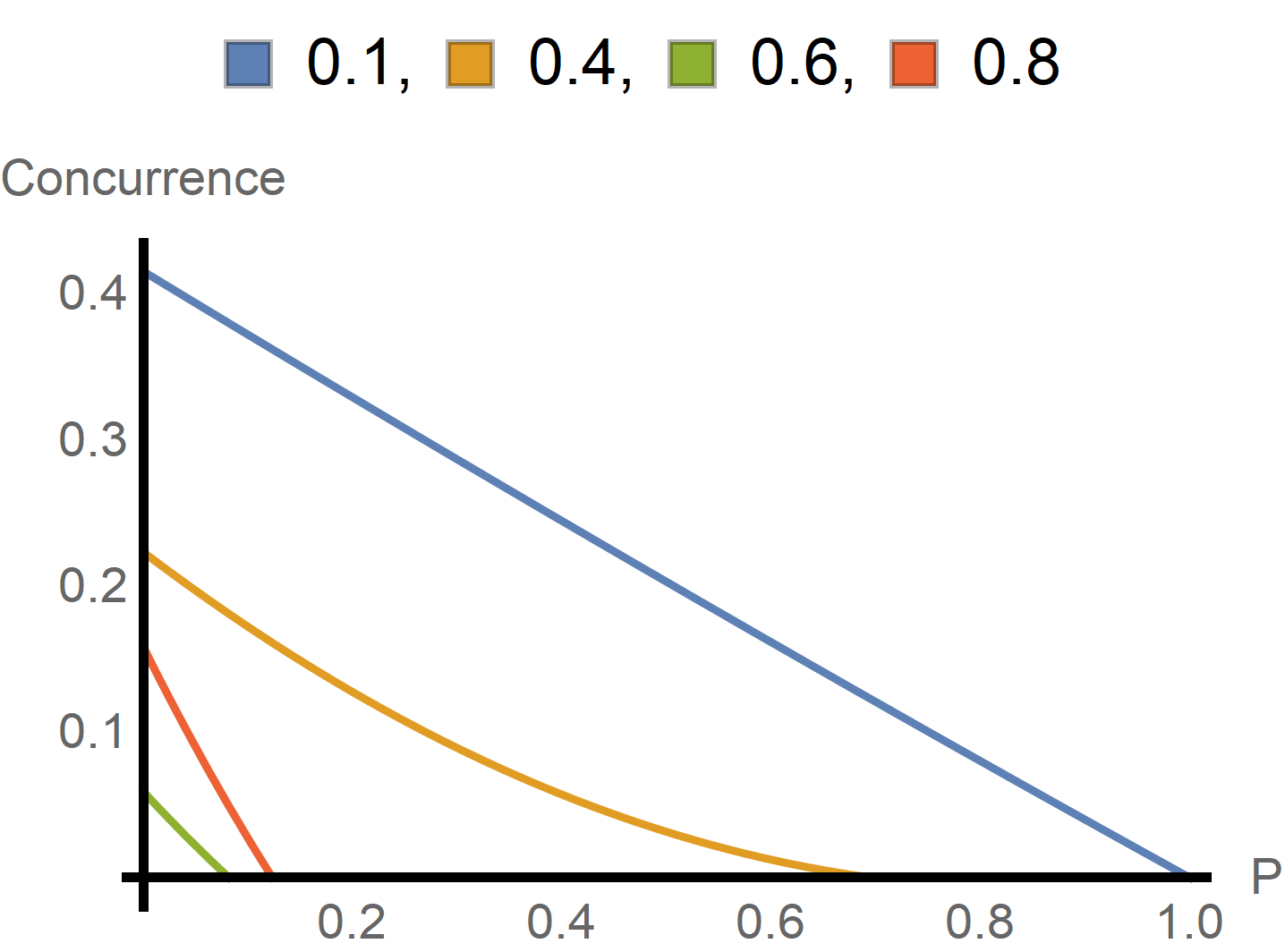}
\caption{In this figure, the concurrence is plotted as a function of $P$ for four different interparticle entangled states with input state parameter $d=0.1,0.4,0.6$ and $0.8$. For these four states, other input state parameters are $a=c=\frac{1}{4}$.} 

\label{fig:amplinter}
\end{figure}
 
Next, we compare the concurrence of intraparticle and interparticle entangled states with the same coefficients $a$, $b$, $c$ and $d$. The results for representative values are shown in Fig.~\ref{fig:ampcomp}. The intraparticle case displays a rebirth of entanglement, while the interparticle case does not, as remarked earlier. Lastly, it can be seen that the concurrence in the interparticle case is lower than in the interparticle case for the same channel strength, demonstrating that the former has suppressed quantum correlations more strongly than the latter.
 \begin{figure}[h]
\centering
\includegraphics[scale=0.8]{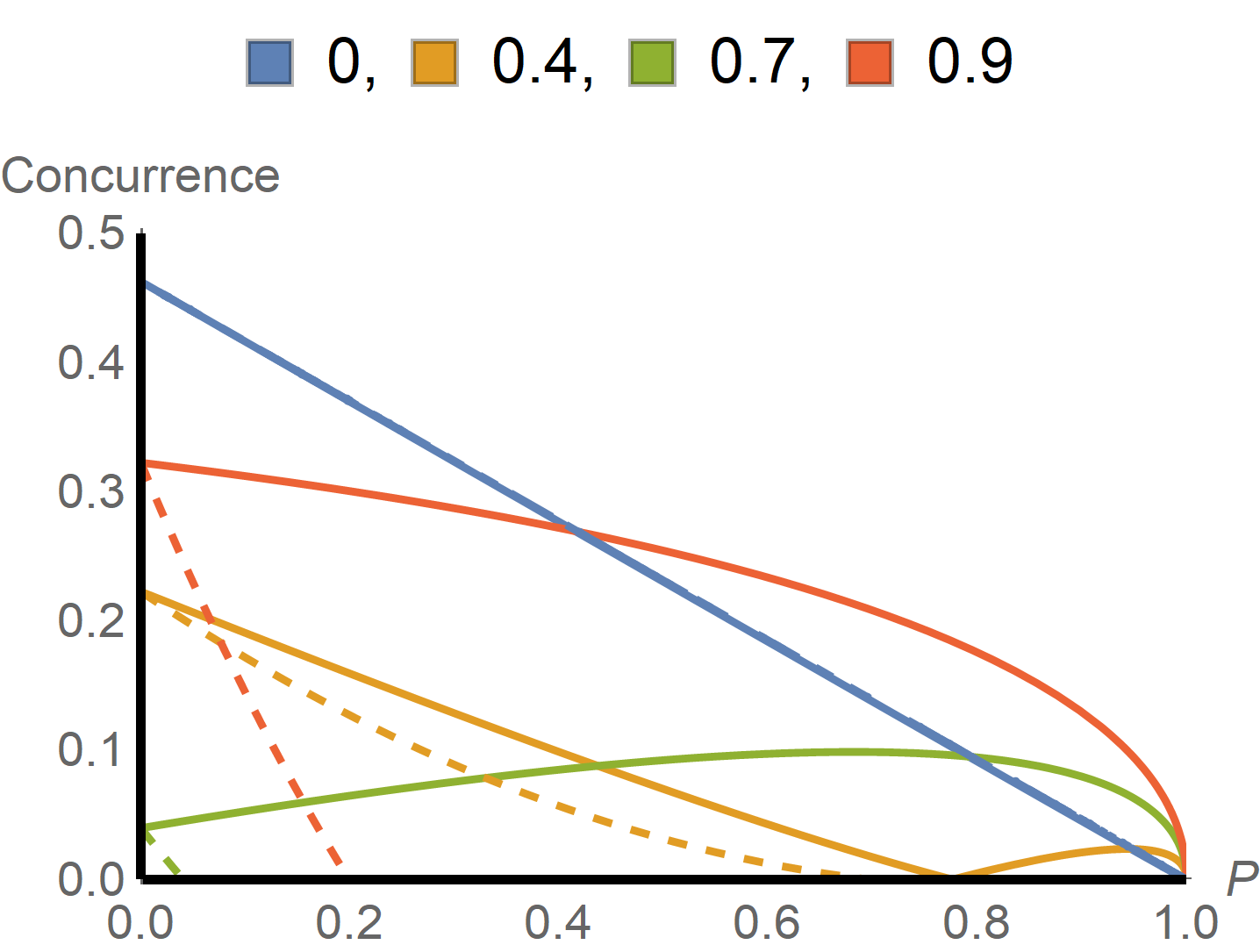}
\caption{In this figure, the concurrence for intraparticle entangled states and interparticle entangled states are plotted as a function of $P$ for four different values of $d=$ $0$, $0.4$, $0.7 $ and $0.9 $. For these four states, other input state parameters are $a=c=\frac{1}{4}$. In the figure, solid curves represent the concurrence for intraparticle entangled states, while dashed curves represent the concurrence for interparticle entangled states. From the curve, it is clear that only intraparticle entanglement demonstrates rebirth and creation of entanglement.}
\label{fig:ampcomp}
\end{figure}

Revival of interparticle entanglement is possible under special circumstances. In \cite{BBellomo}, the authors have shown a revival of interparticle entanglement under the non-Markovian dissipative Amplitude damping channel where the channel parameter takes a special form
\begin{eqnarray}
P = e^{-\Gamma t}\Bigg( \cos (\frac{dt}{2})+\frac{\Gamma}{d}\sin (\frac{dt}{2})\Bigg)^{2}
\end{eqnarray}
where $d = \sqrt{2\gamma\Gamma - \Gamma^{2}}$ and $t$ is the time parameter. Note, however that, unlike the intraparticle case we consider, an initially unentangled state continues to remain unentangled. In order to get a revival of interparticle entanglement one needs the specific model of the channel parameter, while one can get a revival of intraparticle entanglement without considering any specific model of the channel parameter.  Thus, the revival of intraparticle entanglement under the amplitude damping channel is model independent whereas, one can observe the revival of interparticle entanglement only under specific model dependent forms for the amplitude damping channel.

 There are interesting works in the literature \cite{Brun2002,Hamdouni2006,Ficek2006,Sun2007} on the revival of interparticle entanglement under decoherence. If we look carefully, we notice that in those works, the environment is a very large dimensional quantum system. These cases are markedly different from our cases where the dimension of the Hilbert space of the environment is comparable to the Hilbert space dimension of the system.

\section{\label{sec:level6} Other damping channels}
We have investigated the effect of the phase damping and the depolarizing channels separately on a pure intraparticle entangled state $|\psi\rangle_{in}$ described by Eqn.~\ref{eqn:inputstateMain}. The detailed investigation is given in the Appendix \ref{sec:appendix2} and Appendix \ref{sec:appendix3}. We do not observe any rebirth of intraparticle entanglement after entanglement sudden death or creation of intraparticle entanglement under these two damping channels. We have also calculated the decay of concurrence of a pure interparticle entangled state under the phase damping and depolarizing channels and have compared this result to the decay for an intraparticle entangled state. We find that the decay of interparticle entanglement is much faster than the decay of intraparticle entanglement for these channels, which is similar to the case of amplitude damping channel.

\section{\label{sec:level7} Future Scope}
In this study, we have conducted a theoretical examination of the decoherence behavior of a pure, generic intraparticle entangled state. Specifically, we analyzed how this state is affected by three different types of quantum noise channels: amplitude damping, phase damping and depolarizing channels. It is important to note that our analysis is not tied to any particular physical implementation of an intraparticle entangled system. Instead, our findings are broadly applicable to any such system, regardless of the specific physical system details.

The focus of our theoretical analysis is to provide a general understanding of the decoherence mechanisms without being constrained by the peculiarities of a specific system. This generality allows our results to be versatile and widely applicable.


As a future scope, we can consider any physical system in which intraparticle states undergo decoherence and verify the results obtained for different damping channels. Our study aids in understanding the state parameters and in modeling experiments with reduced decoherence effects. This helps to achieve intraparticle entangled states as an efficient resource, enabling researchers to design quantum systems more effectively, with minimal information loss, thereby ensuring the successful completion of quantum information tasks.
\section{\label{sec:level10}Concluding Remarks}
To conclude, we have studied the generation and decay of intraparticle entanglement under the amplitude damping channel by calculating the concurrence. We obtain an exact analytical expression for the concurrence for an arbitrary initial pure state. The concurrence exhibits the interesting phenomenon of entanglement revival as a function of the damping parameter. We also calculate the concurrence as a function of the channel parameter for the more conventional case of interparticle entanglement and show that there is no revival of entanglement or indeed any non-monotonic dependence on the channel parameter. Thus, the intraparticle system displays the counterintuitive phenomenon of the growth of entanglement with increasing damping strength. We provide an explanation for this in terms of the number of non-zero eigenvalues of the output density matrix. Further, the entanglement decay is much greater for the interparticle as compared to the intraparticle case for the same input state and channel parameter. We also study the effects of phase damping and the depolarizing channel for both the intraparticle and interparticle cases and find no evidence of entanglement revival in either. However, the decay of entanglement for comparable parameters is always greater for the interparticle case as compared to the intraparticle case consistent with our observations for the amplitude damping channel. Thus, intraparticle entanglement is in general more robust than interparticle entanglement.

\begin{acknowledgments}
US and PKP acknowledge partial support from the DST-ITPAR grant IMT/Italy/ITPAR-IV/QP/2018/G. US acknowledges partial support provided by the Ministry of Electronics and Information
Technology (MeitY), Government of India under grant for Centre for Excellence in Quantum
Technologies with Ref. No. 4(7)/2020 – ITEA and QuEST-DST project Q-97 of the Govt.of India. SM acknowledges support from QuEST-DST project Q-34 of the Govt.of India. US and PKP acknowledge Dipankar Home and PKP acknowledges Abhinash K Roy for many useful discussions. 
\end{acknowledgments}

\begin{widetext}
\appendix
\section{\label{sec:appendix1}Amplitude Damping Channel}
Our input intraparticle entangled state is
\begin{eqnarray}
\label{eqn:inputstate}
|\psi\rangle_{in} = a|a_{1}b_{1}\rangle +b|a_{1}b_{2}\rangle +c|a_{2}b_{1}\rangle +d|a_{2}b_{2}\rangle = a|a_{1}b_{1}\rangle + \sqrt{1-|a|^{2}}|\phi\rangle
\end{eqnarray}
where $a$, $b$, $c$ and $d$ are the complex input state parameters and
\begin{eqnarray}
\label{eqn:phi}
|\phi\rangle = \frac{1}{\sqrt{1-|a|^{2}}} (b|a_{1}b_{2}\rangle +c|a_{2}b_{1}\rangle +d|a_{2}b_{2}\rangle)
\end{eqnarray}
Here, $|a|$ is the amplitude of the complex input state parameter $a$. Two states $|a_{1}b_{1}\rangle$ and $|\phi\rangle$ form a orthonormal basis: $\langle a_{1}b_{1}|\phi\rangle = 0 ~~and ~~ \langle a_{1}b_{1}|a_{1}b_{1}\rangle = \langle \phi|\phi\rangle = 1 $. So, the state $|\psi\rangle_{in}$ looks like a qubit state.
  
  The Kraus operators representing the amplitude damping channel is \cite{ArijitDutta, AlejandroFonseca}
  \begin{eqnarray*}
  M_{0} = |0\rangle\langle 0| + \sqrt{1-P}\sum_{j=0}^{3} |j\rangle\langle j|~~,~~M_{i} = \sqrt{P}|0\rangle\langle j|~~, ~~j=1,2,3
  \end{eqnarray*}
where $P$ is the channel parameter and the states $|0 \rangle=|a_1b_1\rangle$, $|1 \rangle=|a_1b_2\rangle$, $|2 \rangle=|a_2b_1\rangle$ and $|3 \rangle=|a_2b_2\rangle$ for brevity. The state $|0\rangle$ corresponds to the ground state of the system, and since the damping affects both degrees of freedom simultaneously, we assume that 
it causes transitions from all excited states to the ground state in an equivalent manner. The output state under the effect of the amplitude damping channel is given by
\begin{eqnarray}
\label{eqn:inputstate1}
\rho = \sum_{i=0}^{3}M_{i}\rho_{in}M_{i}^{\dagger} &=& [P+(1-P)|a|^{2}]|a_{1}b_{1}\rangle\langle a_{1}b_{1}| + \sqrt{1-P}\sqrt{1-|a|^{2}}(a|a_{1}b_{1}\rangle\langle \phi |+a^{*}|\phi\rangle\langle a_{1}b_{1}|)\nonumber\\
&+&(1-P)(1-|a|^{2})|\phi\rangle\langle\phi |
\end{eqnarray}
Where $a^{*}$ is the complex conjugate of the input state parameter $a$ and $\rho_{in}=|\psi\rangle_{in}\langle\psi |$ is the input state. In the density matrix form, the output state becomes:
\begin{eqnarray*}
\rho = \begin{pmatrix}
P+(1-P)|a|^{2} & \sqrt{1-P}\sqrt{1-|a|^{2}}a^{*} \\
\sqrt{1-P}\sqrt{1-|a|^{2}}a &(1-P)(1-|a|^{2})
\end{pmatrix}
\end{eqnarray*}
Since $\rho$ is a $2\times 2$ matrix, it has two eigenvalues:
\begin{eqnarray}
\label{eqn:eigen1}
\lambda_{1} = \frac{1}{2} + \alpha ~~,~~\lambda_{2} = \frac{1}{2} - \alpha ~~,~~ \text{where}~~ \alpha = \sqrt{\frac{1}{4} - P(1-P)(1-|a|^{2})^{2}}
\end{eqnarray}
The corresponding eigenvectors are
\begin{eqnarray}
\label{eqn:eigenvalue1}
|u_{1}\rangle = \cos\frac{\theta}{2}|a_{1}b_{1}\rangle +\sin\frac{\theta}{2}e^{-i\xi}|\phi\rangle ~~,~~
|u_{2}\rangle = \sin\frac{\theta}{2}|a_{1}b_{1}\rangle -\cos\frac{\theta}{2}e^{-i\xi}|\phi\rangle
\end{eqnarray}
where
\begin{eqnarray}
\cos\theta = \frac{1}{\alpha}[P+(1-P)|a|^{2}-\frac{1}{2}] ~~,~~ \sin\theta = \frac{1}{\alpha}\sqrt{1-P}\sqrt{1-|a|^{2}}|a|~~and,~~ a = |a|e^{i\xi}
\end{eqnarray}
Now consider, 
\begin{eqnarray}
\rho^{*} = [P+(1-P)|a|^{2}]|a_{1}b_{1}\rangle\langle a_{1}b_{1}| + \sqrt{1-P}\sqrt{1-|a|^{2}}(a^{*}|a_{1}b_{1}\rangle\langle \phi^{*} |+a|\phi^{*}\rangle\langle a_{1}b_{1}|) +(1-P)(1-|a|^{2})|\phi^{*}\rangle\langle\phi^{*} |
\end{eqnarray}
where 
\begin{eqnarray*}
|\phi^{*}\rangle = \frac{1}{\sqrt{1-|a|^{2}}} (b^{*}|a_{1}b_{2}\rangle +c^{*}|a_{2}b_{1}\rangle +d^{*}|a_{2}b_{2}\rangle)
\end{eqnarray*}
Let us define $U=\sigma_{y}\otimes \sigma_{y}$. In the basis $\{ |a_{1}b_{1}\rangle, |a_{1}b_{2}\rangle, |a_{2}b_{1}\rangle, |a_{2}b_{2}\rangle \}$, this has the form
\begin{eqnarray*}
U = \begin{pmatrix}
0 & 0 & 0 & -1 \\
0 & 0 & 1 & 0 \\
0 & 1 & 0 & 0 \\
-1 & 0 & 0 & 0 
\end{pmatrix}
\end{eqnarray*}
$U$ is a unitary matrix. $\rho^{'}= U\rho^{*}U$ is thus a unitary transformation on $\rho^{*}$. This, coupled with the fact that $\rho$ is Hermitian $\implies$ $\rho^{'}$ is also Hermitian with the same eigenvalues as $\rho$. The action of $U$ on the states  $ |a_{1}b_{1}\rangle, |a_{1}b_{2}\rangle, |a_{2}b_{1}\rangle, |a_{2}b_{2}\rangle $ is
\begin{eqnarray*}
U|a_{1}b_{1}\rangle = -|a_{2}b_{2}\rangle ~,~ U|a_{1}b_{2}\rangle = |a_{2}b_{1}\rangle ~,~ U|a_{2}b_{1}\rangle = |a_{1}b_{2}\rangle ~,~ U|a_{2}b_{2}\rangle = -|a_{1}b_{1}\rangle
\end{eqnarray*}
Thus,
\begin{eqnarray}
\label{eqn:inputstate2}
\rho^{'} =  [P+(1-P)|a|^{2}]|a_{2}b_{2}\rangle\langle a_{2}b_{2}| - \sqrt{1-P}\sqrt{1-|a|^{2}}(a^{*}|a_{2}b_{2}\rangle\langle \phi^{'} |+a|\phi^{'}\rangle\langle a_{2}b_{2}|) +(1-P)(1-|a|^{2})|\phi^{'}\rangle\langle\phi^{'} |
\end{eqnarray}
where
\begin{eqnarray}
\label{eqn:phiprime}
|\phi^{'}\rangle = U |\phi^{*}\rangle = \frac{1}{\sqrt{1-|a|^{2}}} (b^{*}|a_{2}b_{1}\rangle +c^{*}|a_{1}b_{2}\rangle - d^{*}|a_{1}b_{1}\rangle)
\end{eqnarray}
It is straightforward to see that $\langle a_{2}b_{2} | \phi^{'}\rangle =0$ and $\langle \phi^{'} | \phi^{'}\rangle = 1$. Comparing Eq.~(\ref{eqn:inputstate1}) and Eq.~(\ref{eqn:inputstate2}), it is clear that $\rho$ and $\rho^{'}$ have the same eigenvalues. By comparison with Eq.~(\ref{eqn:eigenvalue1}), the eigenvectors of $\rho^{'}$ are
\begin{eqnarray}
\label{eqn:eigenvector2}
|v_{1}\rangle = -\cos\frac{\theta}{2}|a_{2}b_{2}\rangle +\sin\frac{\theta}{2}e^{i\xi}|\phi^{'}\rangle ~~,~~
|v_{2}\rangle = -\sin\frac{\theta}{2}|a_{2}b_{2}\rangle -\cos\frac{\theta}{2}e^{i\xi}|\phi^{'}\rangle
\end{eqnarray}

 Note that even though $\{ |u_{1}\rangle , |u_{2}\rangle\}$ and $\{ |v_{1}\rangle , |v_{2}\rangle\}$ are each a set of two orthonormal vectors, there is no $2\times 2$ unitary transformation between them. This is because those vectors are embedded in a higher (four dimensional ) space.

  From Eq.~(\ref{eqn:phi}) and Eq.~(\ref{eqn:phiprime}), we find
  \begin{eqnarray}
\langle a_{1}b_{1}|\phi^{'}\rangle = -\langle \phi |a_{2}b_{2}\rangle = -\frac{d^{*}}{\sqrt{1-|a|^{2}}} ~~\text{and,}~~ \langle \phi |\phi^{'}\rangle = \frac{2b^{*}c^{*}}{1-|a|^{2}}
  \end{eqnarray}
These can be used to obtain
\begin{eqnarray}
\label{eqn:innerproductuv}
\langle u_{1}|v_{1}\rangle &=& -\frac{d^{*}e^{i\xi}\sin\theta}{\sqrt{1-|a|^{2}}} + \frac{b^{*}c^{*}e^{i2\xi}(1-\cos\theta)}{1-|a|^{2}}\nonumber\\
\langle u_{2}|v_{2}\rangle &=& \frac{d^{*}e^{i\xi}\sin\theta}{\sqrt{1-|a|^{2}}} + \frac{b^{*}c^{*}e^{i2\xi}(1+\cos\theta)}{1-|a|^{2}}\nonumber\\
\langle u_{1}|v_{2}\rangle &=& \langle u_{2}|v_{1}\rangle= \frac{d^{*}e^{i\xi}\cos\theta}{\sqrt{1-|a|^{2}}} - \frac{b^{*}c^{*}e^{i2\xi}\sin\theta}{1-|a|^{2}}
\end{eqnarray}

 Now consider the matrix $R = \rho (\sigma_{y}\otimes\sigma_{y})\rho^{*}(\sigma_{y}\otimes\sigma_{y})=\rho U\rho^{*}U = \rho\rho^{'}$, which is used to calculate the concurrence of the output state $\rho$. We have provided the definition of $R$ on page number $2$ of the main manuscript. Note that $R$ is not a Hermitian matrix. Let us write $R$ in the basis of the orthonormal vectors $|v_{1}\rangle$, $|v_{2}\rangle$, $|v_{3}\rangle$ and $|v_{4}\rangle$. $|v_{1}\rangle$, $|v_{2}\rangle$, $|v_{3}\rangle$ and $|v_{4}\rangle$ are the eigenvalues of $\rho^{'}$. $|v_{1}\rangle$ and $|v_{2}\rangle$ are defined in Eq.~(\ref{eqn:eigenvector2}), $|v_{3}\rangle$ and $|v_{4}\rangle$ are the eigenvectors of the subspace corresponding to the two zero eigenvalues of $\rho^{'}$. The exact forms of $|v_{3}\rangle$ and $|v_{4}\rangle$ are unimportant, but we need them to formally write down $R$ as a $4\times 4$ matrix. Thus, $\rho^{'}|v_{1}\rangle = \lambda_{1}|v_{1}\rangle$, $\rho^{'}|v_{2}\rangle = \lambda_{2}|v_{2}\rangle$, $\rho^{'}|v_{3}\rangle =0$ and $\rho^{'}|v_{4}\rangle = 0$. The elements of the matrix $R$ are $R_{ij}=\langle v_{i}|\rho\rho^{'}|v_{j}\rangle$. Thus
\begin{eqnarray*}
R = \begin{pmatrix}
R_{11} & R_{12} & 0 & 0 \\
R_{21} & R_{22} & 0 & 0 \\
R_{31} & R_{32} & 0 & 0 \\
R_{41} & R_{42} & 0 & 0 
\end{pmatrix}
\end{eqnarray*} 
The eigenvalues of $R$ can be obtained from the determinant equation
\begin{eqnarray*}
\begin{vmatrix}
R_{11}-\lambda & R_{12} & 0 & 0 \\
R_{21} & R_{22}-\lambda & 0 & 0 \\
R_{31} & R_{32} & -\lambda & 0 \\
R_{41} & R_{42} & 0 & -\lambda
\end{vmatrix}
=0 \implies \big[ (\lambda - R_{11})(\lambda - R_{22})-R_{12}R_{21}\big]\lambda^{2} = 0
\end{eqnarray*} 
This immediately tells us that two of the eigenvalues of $R$ equal zero. The two nonzero eigenvalues are the roots of the quadratic equation
\begin{eqnarray*}
(\lambda - R_{11})(\lambda - R_{22})-R_{12}R_{21} =0
\end{eqnarray*}
Since $R=\rho\rho^{'}$, $\rho^{'}|v_{1}\rangle = \lambda_{1}|v_{1}\rangle$ and $\rho^{'}|v_{2}\rangle = \lambda_{2}|v_{2}\rangle$, we obtain the equation
\begin{eqnarray*}
(\lambda - \lambda_{1}\rho_{11})(\lambda - \lambda_{2}\rho_{22}) - \lambda_{1}\lambda_{2}\rho_{12}\rho_{21} = 0
\end{eqnarray*}
where,
\begin{eqnarray}
\rho_{ij} = \langle v_{i} |\rho | v_{j}\rangle ~~,~~ \text{for}, i,j \in \{1 , 2 \}
\end{eqnarray}
The two nonzero eigenvalues are thus
\begin{eqnarray}
\label{eqn:eigenvaluesR}
\lambda_{\pm} = \frac{1}{2}\Big[ \lambda_{1}\rho_{11} + \lambda_{2}\rho_{22} \pm \sqrt{(\lambda_{1}\rho_{11} + \lambda_{2}\rho_{22})^{2}-4\lambda_{1}\lambda_{2}(\rho_{11}\rho_{22}-\rho_{12}\rho_{21})} \Big]
\end{eqnarray}
Let us define,
\begin{eqnarray}
\label{eqn:sandt}
S &=& \lambda_{1}\rho_{11} + \lambda_{2}\rho_{22} \nonumber\\
T &=& 2\sqrt{\lambda_{1}\lambda_{2}(\rho_{11}\rho_{22}-\rho_{12}\rho_{21})}
\end{eqnarray}
Then
\begin{eqnarray}
\lambda_{\pm} = \frac{S\pm \sqrt{S^{2}-T^{2}}}{2}
\end{eqnarray}
The concurrence is
\begin{eqnarray}
\label{eqn:expression-c}
C = \sqrt{\lambda_{+}} - \sqrt{\lambda_{-}} \implies C^{2} = \lambda_{+} + \lambda_{-} -2\sqrt{\lambda_{+} \lambda_{-}} = S - T
\end{eqnarray}
It only remains to evaluate $S$ and $T$ to obtain the concurrence. Now,
\begin{eqnarray*}
S=\lambda_{1}\rho_{11} + \lambda_{2}\rho_{22}
\end{eqnarray*}
where $\rho_{ij} = \langle v_{i} |\rho | v_{j}\rangle$, but $\rho = \lambda_{1}|u_{1}\rangle\langle u_{1}| + \lambda_{2}|u_{2}\rangle\langle u_{2}|$ with $u_{1}$, $u_{2}$ given by Eq.~(\ref{eqn:eigenvalue1}). Substituting $\rho$ into the $\rho_{ij}$ we get
\begin{eqnarray}
\label{eqn:rho11}
\rho_{11} &=& \langle v_{1} |\rho | v_{1}\rangle = \lambda_{1}|U_{11}|^{2} + \lambda_{2}|U_{12}|^{2}\nonumber\\
\rho_{22} &=& \langle v_{2} |\rho | v_{2}\rangle = \lambda_{1}|U_{21}|^{2} + \lambda_{2}|U_{22}|^{2}
\end{eqnarray}
where $U_{ij}=\langle u_{i}|v_{j}\rangle$ for $i, j \in \{ 1, 2\}$. Thus
\begin{eqnarray*}
S = \lambda_{1}^{2}|U_{11}|^{2} + \lambda_{2}^{2}|U_{22}|^{2} + \lambda_{1}\lambda_{2}(|U_{12}|^{2} + |U_{21}|^{2})
\end{eqnarray*}
Writing $\lambda_{1}$ and $\lambda_{2}$ in terms of $\alpha$ from Eq.~(\ref{eqn:eigen1}),
\begin{eqnarray}
\label{eqn:s1}
S = \frac{1}{4}\big[|U_{11}|^{2} + |U_{22}|^{2} + |U_{12}|^{2} + |U_{21}|^{2}  \big] + \alpha^{2}\big[|U_{11}|^{2} + |U_{22}|^{2} - |U_{12}|^{2} - |U_{21}|^{2}  \big] +\alpha\big[|U_{11}|^{2} - |U_{22}|^{2} \big]
\end{eqnarray}
From Eq.~(\ref{eqn:innerproductuv}),
\begin{eqnarray}
|U_{11}|^{2} &=& \frac{|b|^{2}|c|^{2}(1-\cos\theta)^{2}}{(1-|a|^{2})^{2}} + \frac{|d|^{2}\sin^{2}\theta}{1-|a|^{2}} - \frac{(1-\cos\theta)\sin\theta (b^{*}c^{*}de^{i\xi} + bcd^{*}e^{-i\xi})}{(1-|a|^{2})^{\frac{3}{2}}}\nonumber\\
|U_{22}|^{2} &=&  \frac{(1+\cos\theta)^{2}}{(1-|a|^{2})^{2}} +\frac{|d|^{2}\sin^{2}\theta}{1-|a|^{2}} + \frac{(1+\cos\theta)\sin\theta (b^{*}c^{*}de^{i\xi} + bcd^{*}e^{-i\xi})}{(1-|a|^{2})^{\frac{3}{2}}}\nonumber\\
|U_{12}|^{2} &=& |U_{21}|^{2} = \frac{|b|^{2}|c|^{2}\sin^{2}\theta}{(1-|a|^{2})^{2}} + \frac{|d|^{2}\cos^{2}\theta}{1-|a|^{2}} - \frac{\sin\theta\cos\theta (b^{*}c^{*}de^{i\xi} + bcd^{*}e^{-i\xi})}{(1-|a|^{2})^{\frac{3}{2}}}
\end{eqnarray}
Substituting these expressions of $|U_{11}|^{2}$, $|U_{12}|^{2}=|U_{21}|^{2}$, $|U_{22}|^{2}$ into the Eq.~(\ref{eqn:s1}) and simplifying we get
\begin{eqnarray}
\label{eqn:expression-s}
S = 4|b|^{2}|c|^{2}(1-P)^{2} - 4(1-P)^{3/2}\big[adb^{*}c^{*} + bca^{*}d^{*}] + 2|d|^{2}(1-P) [P + (2 - P)|a|^{2}]
\end{eqnarray}
From Eq.~(\ref{eqn:sandt}), $T=2\sqrt{\lambda_{1}\lambda_{2}(\rho_{11}\rho_{22}-\rho_{12}\rho_{21})}$. we have already calculated $\rho_{11}$ and $\rho_{22}$ in Eq.~(\ref{eqn:rho11}). In the similar way we can also calculate $\rho_{12}$ and $\rho_{21}$:
\begin{eqnarray*}
\rho_{12} &=& \langle v_{1} |\rho | v_{2}\rangle = \lambda_{1}\langle v_{1}|u_{1}\rangle\langle u_{1}|v_{2}\rangle + \lambda_{2}\langle v_{1}|u_{2}\rangle\langle u_{2}|v_{2}\rangle\nonumber\\
\rho_{21} &=& \langle v_{2} |\rho | v_{1}\rangle = \lambda_{1}\langle v_{2}|u_{1}\rangle\langle u_{1}|v_{1}\rangle + \lambda_{2}\langle v_{2}|u_{2}\rangle\langle u_{2}|v_{1}\rangle
\end{eqnarray*}
Then
\begin{eqnarray*}
\rho_{11}\rho_{22} - \rho_{12}\rho_{21} = \lambda_{1}\lambda_{2}|\langle u_{1}|v_{1}\rangle\langle u_{2}|v_{2}\rangle - \langle u_{2}|v_{1}\rangle\langle u_{1}|v_{2}|^{2}
\end{eqnarray*}
Substituting the above expression of $\rho_{11}\rho_{22} - \rho_{12}\rho_{21}$ into the expression of $T$ (Eq.~(\ref{eqn:sandt})) and using Eq.~(\ref{eqn:innerproductuv}) we get,
\begin{eqnarray}
\label{eqn:expression-t}
T = 2P(1-P)(1-|a|^{2})|d|^{2}
\end{eqnarray}
From Eq.~(\ref{eqn:expression-c}), Eq.~(\ref{eqn:expression-s}) and Eq.~(\ref{eqn:expression-t}) we have
\begin{eqnarray*}
C^{2} = S-T = 4|bc\sqrt{1-P} - ad|^{2}(1-P)
\end{eqnarray*}
So, the expression of the concurrence becomes
\begin{eqnarray}
C = 2|bc\sqrt{1-P} - ad|\sqrt{1-P}
\end{eqnarray}
This is the exact analytical expression of the concurrence of the output intraparticle entangled state (Eq.~(\ref{eqn:inputstate1})) with complex input state parameters $a$, $b$, $c$ and $d$ under the effect of the amplitude damping channel. To get a better understanding of the concurrence, let us consider the input state parameters as
\begin{eqnarray}
a = |a|e^{i\theta_{a}}~~,~~b = |b|e^{i\theta_{b}}~~,~~c = |c|e^{i\theta_{c}}~~,~~d = |d|e^{i\theta_{d}}
\end{eqnarray}
Substituting these parameters, we get
\begin{eqnarray*}
C = 2\Big||b||c|\sqrt{1-P}e^{i(\theta_{b}+\theta_{c})}-|a||d|e^{i(\theta_{a}+\theta_{d})}\Big|\sqrt{1-P}
\end{eqnarray*}
Simplifying the above equation, we get
\begin{eqnarray*}
C = 2\Big[ |b|^{2}|c|^{2}(1-P)+|a|^{2}|d|^{2}-2|a||b||c||d|\sqrt{1-P}\cos (\Delta\theta) \Big]^{1/2} \sqrt{1-P}
\end{eqnarray*}
where $\Delta\theta = \theta_{b}+\theta_{c}-\theta_{a}-\theta_{d}$. It is interesting to note that the concurrence depends on the difference between $\theta_{b}+\theta_{c}$ and $\theta_{a}+\theta_{d}$. If an individual angle changes, keeping the difference fixed, the concurrence also remains fixed. Let us now try some geometric representations of the concurrence. Let us suppose $\vec{V}_{1}$ and $\vec{V}_{2}$ are two vectors given by
\begin{eqnarray*}
\vec{V}_{1} = |b||c|\sqrt{1-P}e^{i(\theta_{b}+\theta_{c})} ~~,~~\vec{V}_{2} = |a||d|e^{i(\theta_{a}+\theta_{d})}
\end{eqnarray*}
Then, the expression of concurrence is given by
\begin{eqnarray*}
C = 2|\vec{V}_{1} - \vec{V}_{2}|\sqrt{1-P} = 2|\vec{V}|\sqrt{1-P}
\end{eqnarray*}
where $\vec{V}=\vec{V}_{1} - \vec{V}_{2}$. So, the concurrence is the magnitude of the vector $\vec{V}$ and then multiplied by  $2\sqrt{1-P}$.
\begin{figure}[h]
\centering
\includegraphics[scale=1]{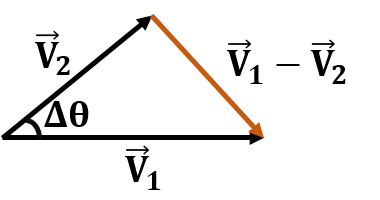}
\caption{In this figure $\vec{V}_{1}$, $\vec{V}_{2}$, $\vec{V}=\vec{V}_{1} - \vec{V}_{2}$ and $\Delta\theta $ are shown.}
\end{figure}
 When the magnitude of the vector $\vec{V}$ is zero, then the concurrence of the intraparticle entangled state also becomes zero. This happens when
\begin{eqnarray}
\label{eqn:concurrence_1}
  |b||c|\sqrt{1-P} = |a||d| ~~,~~\text{and}~~~~  \Delta\theta =0 , \pm 2\pi
\end{eqnarray}
 Eq.~(\ref{eqn:concurrence_1}) implies that
 \begin{eqnarray}
 \label{eqn:concurrence_2}
 |b||c|\sqrt{1-P} = |a||d| \Longrightarrow P = 1 - \Big(\frac{|a||d|}{|b||c|}\Big)^{2}
 \end{eqnarray}
  The condition $\Delta\theta =0 , \pm 2\pi$ arises since the individual angles ($\theta_a$, $\theta_b$, $\theta_c$, $\theta_d$) $\in [0,2\pi]$. As long as $\Delta\theta =0, \pm 2\pi$ and $|a||d| \leq |b||c|$, i.e., $P \leq 1$, we get a physical value of $P$ where entanglement sudden death occurs. When $|a||d| > |b||c|$, $P >1 $ which is unphysical. So, if $|a||d| > |b||c|$, we will not see entanglement sudden death. The variation of concurrence with the channel parameter $P$ for three different input states with state parameters satisfying the condition (i) $|a||d| < |b||c|$, (ii) $|a||d| > |b||c|$ and (iii) $|a||d| = |b||c|$ are represented by the blue, yellow and the green plots in the Fig.~\ref{fig:intra1}. The blue plot shows that if a state satisfies condition (i), we observe a rebirth of entanglement after ESD, which finally asymptotically decays to zero. The yellow plot shows that if a state satisfies condition (ii), we get the initial entanglement creation, which finally decays asymptotically. The green plot shows that if a state satisfies condition (iii), the separable state first becomes an entangled state, and the amount of entanglement initially increases and finally asymptotically decays.
   
   It is important to note that for fixed values of the coefficients $a$, $b$, $c$ and $d$, the concurrence can go to zero only at a particular value of $P$ instead of a range of values. This is a consequence of the fact that two of the eigenvalues of the final state density matrix $\rho$ are equal to zero, and hence, so are two of the eigenvalues of the matrix $R$. The concurrence is calculated using the only two non-zero eigenvalues of the matrix R, as defined in Eq.~(\ref{eqn:eigenvaluesR}). A zero concurrence value results when these two non-zero eigenvalues of $R$ degenerate. This corresponds to a level crossing, which can occur only at isolated points in parameter space, and thus, the concurrence cannot be zero over a range of values of $P$. This behaviour is unique to the intraparticle amplitude damping channel and is in sharp contrast to that observed in the other channels, as shown later. For the other channels, all four eigenvalues of $R$ are generally non-zero; thus, the concurrence can remain zero over a range of values of $P$ without exhibiting a revival.
\begin{figure}[h]
\centering
\includegraphics[scale=0.6]{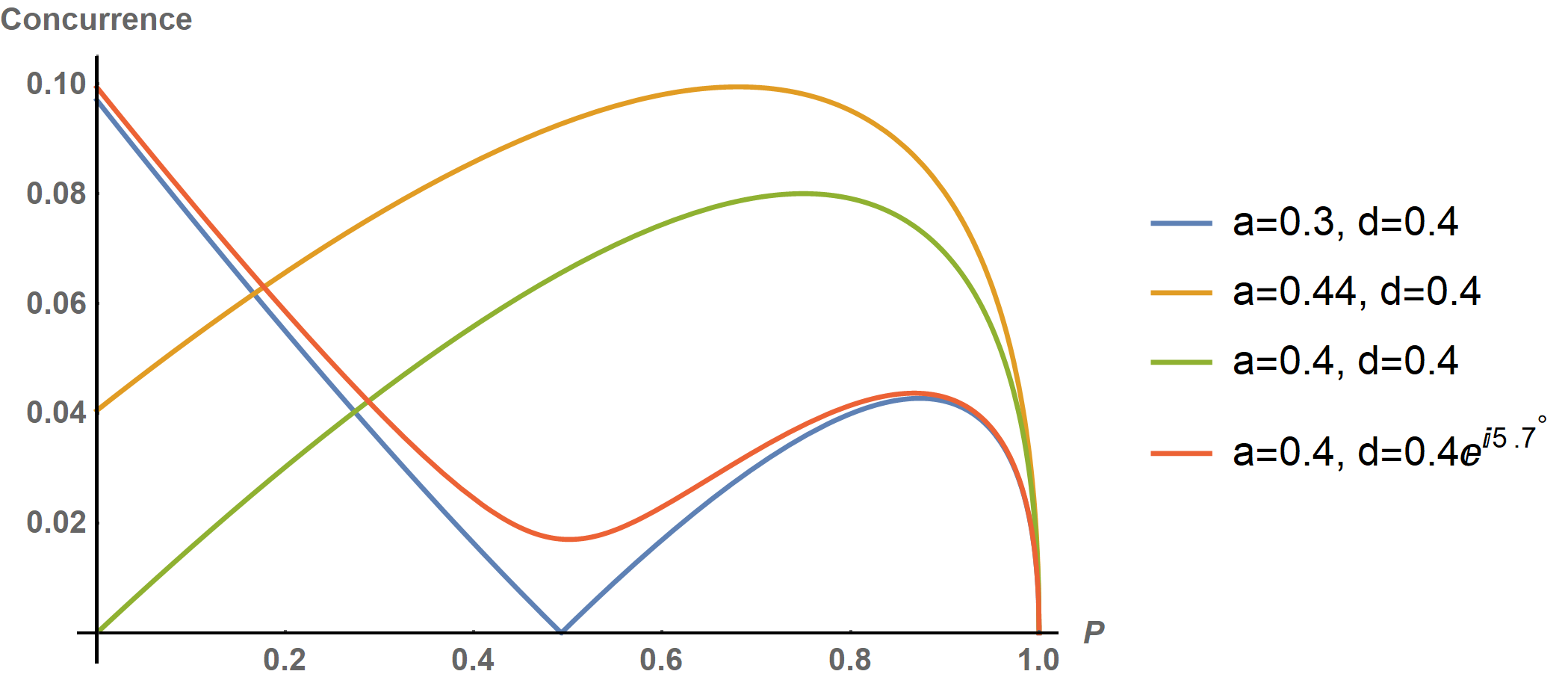}
\caption{The variation of the concurrence with $P$ for three different input states with state parameters $|a|=0.3$, $0.44$ and $0.4$ are represented by the blue, yellow and green plots. Other state parameters for these three input states are $|c|=0.2$,  $|d|=0.4$ and $\Delta\theta = 0$. The variation of Concurrence with channel parameter $P$ for an input state with state parameters $a=0.3$, $b=0.84$, $c=0.2$ and $d=0.4e^{i5.7^{\circ}}$ is represented by the red plot. The state representing the blue plot satisfies the condition (i) $|a||d|<|b||c|$, which first shows entanglement sudden death and then a rebirth of entanglement and finally asymptotically decays to zero. The state representing the yellow plot satisfies the condition (ii) $|a||d|>|b||c|$, which first shows the creation of entanglement and finally asymptotically decays to zero. The state representing the green plot satisfies the condition (iii) $|a||d|>|b||c|$ where the concurrence first becomes nonzero and gradually increases and finally asymptotically decreases to zero. In all three above cases, $\Delta\theta = 0$. If we make $\theta_{d} = 5.7^{\circ}$ while $\theta_{z}=\theta_{b}=\theta_{c}=0$, i.e., $\Delta\theta \neq 0, \pm 2\pi$, then we observe that the concurrence reaches its minimum nonzero value $C_{-}$ and then a maximum values $C_{+}$ and finally asymptotically decays to zero.}
\label{fig:intra1}
\end{figure}
 
  Let us suppose we only consider the condition $\Delta\theta =0, \pm 2\pi$, i.e., even if $\theta_{a}$, $\theta_{b}$, $\theta_{c}$, $\theta_{d}$ are nonzero, they satisfy the relation $\Delta\theta =0, \pm 2\pi$. In that case, the expression of the concurrence becomes
 \begin{eqnarray}
 C &=& 2\Big[ |b|^{2}|c|^{2}(1-P)+|a|^{2}|d|^{2}-2|a||b||c||d|\sqrt{1-P} \Big]^{1/2} \sqrt{1-P}\nonumber\\
 &=& 2\big|\big[|b||c|\sqrt{1-P}-|a||d|\big]\big|\sqrt{1-P}
 \end{eqnarray}
 Next, we consider the case when $\Delta\theta \neq 0, \pm 2\pi$. Let $\sqrt{1-P}=x$, $|b||c|=\alpha$, $|a||d|=\beta$ and the expression of concurrence becomes
   \begin{eqnarray}
   \label{eqn:concurrence_3}
   C = 2\Big[ \alpha^{2}x^{2}+\beta^{2}-2\alpha\beta x\cos (\Delta\theta) \Big]^{1/2} x
   \end{eqnarray}
   So,
   \begin{eqnarray*}
    C^{2} = 4\Big[ \alpha^{2}x^{2}+\beta^{2}-2\alpha\beta x\cos (\Delta\theta) \Big] x^{2}
   \end{eqnarray*}
   For a given set of values of $\alpha$, $\beta$ and $\Delta\theta$ (i.e., for given set of values of $|a|$, $|b|$, $|c|$ and $|d|$), the concurrence has a minimum value when
   \begin{eqnarray*}
   \frac{dC^{2}}{dx} = 8x\big[ 2\alpha^{2}x^{2}+\beta^{2}-3\alpha\beta x\cos \Delta\theta \big] = 0
   \end{eqnarray*}
   Since we are not interested the case $x = 0$, (because $x=0 \Rightarrow P=1$, where concurrence is always zero.)\\
   \begin{eqnarray}
   \label{eqn:concurrence_4}
   \frac{dC^{2}}{dx} = 0 \Rightarrow 2\alpha^{2}x^{2}+\beta^{2}-3\alpha\beta x\cos \Delta\theta = 0
   \end{eqnarray}
   Solving the above equation, we get the solution of $x$ as
   \begin{eqnarray*}
   x = \frac{\beta}{4\alpha}[3\cos\Delta\theta \pm \sqrt{(3\cos\Delta\theta)^{2}-8}]
   \end{eqnarray*}
   Since $\alpha$ and $\beta$ are both positive, for $x$ to be positive, $\cos\Delta\theta$ must be positive, and for $x$ to be real, $\cos^{2}\Delta\theta \geq \frac{8}{9}$, i.e., $\cos\Delta\theta \geq \frac{2\sqrt{2}}{3}$. So, $\Delta\theta$ must lie in the range :
   \begin{eqnarray}
   \label{eqn:deltathetarange}
    0 < \Delta\theta  \leq 19.47~^{\circ} ~~\text{or}~~ 340.53~^{\circ} < \Delta\theta  \leq 360~^{\circ}
   \end{eqnarray}
   In this range of $\Delta\theta$, the acceptable solutions of $x$ are
   \begin{eqnarray}
   \label{eqn:concurrence_5}
   x = \frac{\beta}{4\alpha}[3\cos\Delta\theta \pm \sqrt{(3\cos\Delta\theta)^{2}-8}]
   \end{eqnarray}
   When the concurrence is maximum or minimum, Eq.~(\ref{eqn:concurrence_4}) is satisfied. Substituting Eq.~(\ref{eqn:concurrence_4}) into Eq.~(\ref{eqn:concurrence_3}) we can write the optimum value of the concurrence as
   \begin{eqnarray}
   C_{\mp} = \frac{2}{\sqrt{3}}\big[ \beta^{2}-\alpha^{2}x^{2}\big]^{1/2}x
   \end{eqnarray}
   To verify whether this value of $x$ is indeed making the concurrence minimum or maximum, we consider
   \begin{eqnarray*}
   \frac{d^{2}C^{2}}{dx^{2}} = 48\alpha^{2}x^{2}+8\beta^{2}-48\alpha\beta x\cos \Delta\theta
\end{eqnarray*}
Using Eq.~(\ref{eqn:concurrence_4}) we can write
  \begin{eqnarray*}
   \frac{d^{2}C^{2}}{dx^{2}} = 16\alpha^{2}x^{2}-8\beta^{2}
\end{eqnarray*}
For the concurrence to be minimum,
\begin{eqnarray}
\label{eqn:concurrence_6}
 \frac{d^{2}C^{2}}{dx^{2}} > 0 \Longrightarrow x > \frac{\beta}{\sqrt{2}\alpha}
\end{eqnarray}
To satisfy Eq.~(\ref{eqn:concurrence_6}), the expression of $x$ must satisfy
\begin{eqnarray*}
&&\frac{\beta}{4\alpha}\big[3\cos\Delta\theta \pm \sqrt{(3\cos\Delta\theta)^{2}-8}\big] > \frac{\beta}{\sqrt{2}\alpha}\nonumber\\
&& \Rightarrow \big[3\cos\Delta\theta \pm \sqrt{(3\cos\Delta\theta)^{2}-8}\big] > 2\sqrt{2}
\end{eqnarray*}
We can easily verify that only the left side with the $+$ sign will satisfy this. So, the only acceptable solution of $x$ for which the concurrence becomes minimum is
   \begin{eqnarray}
   \label{eqn:concurrence_7}
   x_{-} = \frac{\beta}{4\alpha}\Big[3\cos\Delta\theta + \sqrt{(3\cos\Delta\theta)^{2}-8}\Big]
   \end{eqnarray}
   and the corresponding value of the channel parameter $P$ where the concurrence takes its minimum value is
   \begin{eqnarray}
   \label{eqn:concurrence_8}
   P_{-} = 1-\frac{\beta^{2}}{16\alpha^{2}}\Big[3\cos\Delta\theta + \sqrt{(3\cos\Delta\theta)^{2}-8}\Big]^{2}
   \end{eqnarray}
   Substituting the expression of $x_{-}$ in to Eq.~(\ref{eqn:concurrence_3}) we get the minimum value of the concurrence:
   \begin{eqnarray}
   \label{eqn:concurrence_9}
   C_{-} = \frac{\beta^{2}}{2\sqrt{3}\alpha}\Big[ 1-\frac{(3\cos\Delta\theta + \sqrt{(3\cos\Delta\theta)^{2}-8})^{2}}{16}\Big]^{1/2}\Big[3\cos\Delta\theta + \sqrt{(3\cos\Delta\theta)^{2}-8} \Big]
   \end{eqnarray}
   When $\Delta\theta = 0, \pm 2\pi$, then
   \begin{eqnarray}
   C_{-}^{\Delta\theta=0, \pm 2\pi} = 0
   \end{eqnarray}

   On the other hand, for the concurrence to be maximum,
\begin{eqnarray}
\label{eqn:concurrence_10}
 \frac{d^{2}C^{2}}{dx^{2}} < 0 \Longrightarrow x < \frac{\beta}{\sqrt{2}\alpha}
\end{eqnarray}
So, the expression of $x$, which satisfies the above condition, is given by
   \begin{eqnarray}
   \label{eqn:concurrence_11}
   x_{+} = \frac{\beta}{4\alpha}\Big[3\cos\Delta\theta - \sqrt{(3\cos\Delta\theta)^{2}-8}\Big]
   \end{eqnarray}
    Thus, the above value of $x$ makes the concurrence maximum, and the corresponding value of $P$ is
     \begin{eqnarray}
     \label{eqn:concurrence_12}
   P_{+} = 1-\frac{\beta^{2}}{16\alpha^{2}}\Big[3\cos\Delta\theta - \sqrt{(3\cos\Delta\theta)^{2}-8}\Big]^{2}
   \end{eqnarray} 
   For a given input intraparticle entangled state, $P_{-}$ is lower than $P_{+}$.
   Substituting the expression of $x_{+}$ in to  Eq.~(\ref{eqn:concurrence_3}) we get the maximum value of the concurrence:
   \begin{eqnarray}
   \label{eqn:concurrence_13}
   C_{+} = \frac{\beta^{2}}{2\sqrt{3}\alpha}\Big[ 1-\frac{(3\cos\Delta\theta - \sqrt{(3\cos\Delta\theta)^{2}-8})^{2}}{16}\Big]^{1/2}\Big[3\cos\Delta\theta - \sqrt{(3\cos\Delta\theta)^{2}-8} \Big]
   \end{eqnarray}
   When $\Delta\theta = 0, \pm 2\pi$, then
   \begin{eqnarray}
   C_{+}^{\Delta\theta=0, \pm 2\pi} = \frac{\beta^{2}}{2\alpha}
   \end{eqnarray}
   To estimate the amount of the creation of entanglement due to the amplitude damping channel, let us define a quantity $\Tilde{C}=\frac{C_{+}-C_{-}}{C_{+}+C_{-}}$. It is clear that $\Tilde{C}$ is only a function of $\Delta\theta$. The behaviour of $\Tilde{C}$ with $\Delta\theta$ is shown in Fig.~\ref{fig:ctilde}. When $\Delta\theta =0~^{\circ}$, $\Tilde{C}=1$. With the increase of $\Delta\theta$, $\Tilde{C}$ decreases gradually and becomes zero at $\Delta\theta=\cos^{-1}{\frac{2\sqrt{2}}{3}}=19.47~^{\circ}$. We have already seen this value of $\Delta\theta$ in Eq.~(\ref{eqn:deltathetarange}). Eq.~(\ref{eqn:deltathetarange}) clearly says that in between $19.47~^{\circ}$ and $340.53~^{\circ}$, $\Delta\theta$ has no physically acceptable value. At $\Delta\theta=340.53~^{\circ}$, $\Tilde{C} = 0$. If we increase $\Delta\theta$ further, $\Tilde{C}$ increases gradually and attains its maximum value $\Tilde{C}=1$ at $\Delta\theta=360~^{\circ}$. So, it is clear that $\Tilde{C}$ is maximum at $\Delta\theta =0, \pm 2\pi$. The above analysis indicates that the maximum amount of rebirth of entanglement occurs when $\Delta\theta =0, \pm 2\pi$. When, $\Delta\theta =0, \pm 2\pi$, the maximum amount of rebirth of entanglement is given by
   \begin{eqnarray}
    C_{creation} = C_{+}^{\Delta\theta=0, \pm 2\pi} - C_{+}^{\Delta\theta=0, \pm 2\pi} = \frac{\beta^{2}}{2\pi} = \frac{|a|^{2}|d|^{2}}{2\pi}.
   \end{eqnarray}
\begin{figure}[h]
\label{fig:ctilde}
\centering
\includegraphics[scale=0.8]{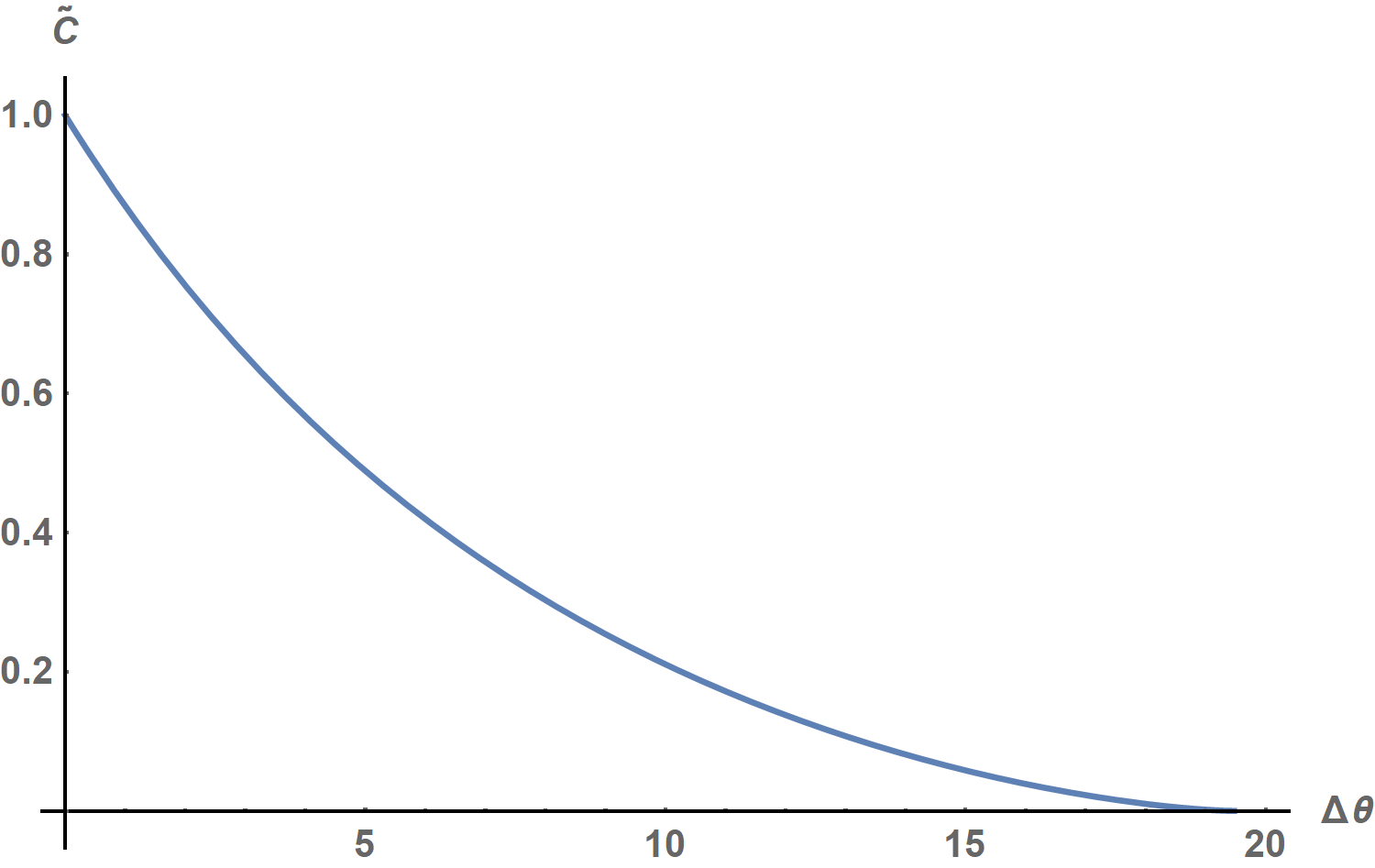}
\caption{Variation of $\Tilde{C}$ with $\Delta\theta$ under the effect of the amplitude damping channel. Here $\Tilde{C}$ is maximum at $\Delta\theta=0$. With the increase of $\Delta\theta$, $\Tilde{C}$ gradually decreases and becomes zero at $\Delta\theta = 19.47^{\circ}$.}
\end{figure}

\section{\label{sec:appendix2}Phase damping Channel}
When a quantum system loses quantum information without losing its energy during the interaction with the environment, the process is known as phase damping. During phase damping, the relative phase among the eigenstates of the system is lost. The Kraus operators for the phase damping channel are:-
\begin{eqnarray}
M_{0} =  \sqrt{1-P}\sum_{j=0}^{3}|j\rangle\langle j| ~~~~,~~~~ M_{i+1} = \sqrt{P}|i\rangle\langle i|,(i=0,1,2,3)
\end{eqnarray}
First we consider an input intraparticle entangled state $|\psi\rangle_{in}$ where input state parameters $a$, $b$, $c$ and $d$ are all real. After evolution of $|\psi\rangle_{in}$ through a phase damping channel, the final state becomes
\begin{eqnarray}
\rho_{out}  = 
\begin{pmatrix}
  a^{2} & ab(1-P) & ac(1-P) & ad(1-P)\\
  ab(1-P) & b^{2} & bc(1-P) & bd(1-P)\\
   ac(1-P) & bc(1-P) & c^{2} & cd(1-P)\\
  ad(1-P) & bd(1-P) & cd(1-P) & d^{2}
\end{pmatrix}
\end{eqnarray}

  Next, we calculate the matrix $R$ and find out all four eigenvalues of this matrix. We have provided the definition of $R$ on page number $2$ of the main manuscript. Eigenvalues of $R$ are
\begin{eqnarray}
\lambda_{1} &=& \frac{1}{2}\Big[\alpha +|(ad-bc)|(2-P)\sqrt{\beta}\Big]~~,~~\lambda_{2} = \frac{1}{2}\Big[\alpha -|(ad-bc)|(2-P)\sqrt{\beta}\Big]~~, ~~\lambda_{3} = a^{2}d^{2}P^{2} ~~,~~\lambda_{4} = b^{2}c^{2}P^{2}
\end{eqnarray}
where
\begin{eqnarray}
\alpha &=& (ad-bc)^{2}(2-P)^{2}+2abcdP(4-3P) ~~,~~\text{and}~~~~\beta = (ad-bc)^{2}(2-P)^{2}+4abcdP(4-3P)
\end{eqnarray} 
Let us consider the situation where $P=0$, i.e. with no effect of phase damping noise. Then, the eigenvalues become
\begin{eqnarray*}
 \lambda_{1} = 4(ad-bc)^{2} ~~, ~~ \lambda_{2}=\lambda_{3}=\lambda_{4}=0
\end{eqnarray*}
So, only $\lambda_{1}$ contributes to the expression of the concurrence, i.e., $C = \sqrt{\lambda_{1}}=2|(ad-bc)|$, while the other three eigenvalues have no contribution to the concurrence. If we increase the channel parameter $P$ gradually, we observe that the eigenvalue $\lambda_{1}$ remains the largest eigenvalue with the increase of $P$ while the other three eigenvalues first become nonzero and then increase gradually. As a result, the concurrence gradually decreases with the channel parameter $P$ increase. Let us now find the channel parameter $P$ value where the concurrence becomes zero. When $C=0$, the following condition is satisfied:
\begin{eqnarray*}
\sqrt{\lambda_{1}} = \sqrt{\lambda_{2}}+\sqrt{\lambda_{3}}+\sqrt{\lambda_{4}}
\end{eqnarray*}
Substituting expressions of all $\lambda_{i}$ in the above equation and solving it for $P$ we get following acceptable solution of $P$:
\begin{equation}
  P =
    \begin{cases}
     ~ \frac{ad-bc}{ad} & ~~\text{if}~~ ad > bc\\
      ~ \frac{bc - ad}{bc} & ~~\text{if}~~ bc > ad
    \end{cases}       
\end{equation}
 For $ad > bc$, if $bc=0$, we only observe asymptotic decay of entanglement. Similarly, for $bc > ad$, if $ad=0$, we only observe asymptotic decay of entanglement. So, in the case of the phase damping channel, we are getting ESD for those input states for which $ad$ and $bc$ are both nonzero. That means that when all input state parameters are nonzero, then only ESD will occur. If one or more input state parameters are zero, we only observe asymptotic decay of entanglement.

From this analysis, we observe that $\lambda_{1}$ is the largest eigenvalue for all input state parameters $a$, $b$, $c$ and $d$, and it remains largest for the entire range of $P$. Let us denote the square root of the largest eigenvalue as $\lambda_{max}$ and the sum of square root of remaining three eigenvalues as $\lambda_{rest}$ : $\lambda_{max}=\sqrt{\lambda_{1}}$, $\sqrt{\lambda_{2}}+\sqrt{\lambda_{3}}+\sqrt{\lambda_{4}} = \lambda_{rest}$. Depending on the values of the input state parameters, when all the input state parameters are nonzero, there is a physical value of the channel parameter $P$ where we observe ESD. Below the ESD point where $\lambda_{max} > \lambda_{rest}$, we get nonzero Concurrence and above the ESD point where $\lambda_{max} < \lambda_{rest}$, we get zero Concurrence. If we observe the condition $\lambda_{max} > \lambda_{rest}$ after the ESD point, we can say that the rebirth of intraparticle entanglement occurs. But, for the phase damping channel, this does not happen.

    In the above analysis, we have considered an intraparticle entangled state whose state parameters $a$, $b$, $c$ and $d$ are all real. Let us now consider an input intraparticle entangled state whose state parameters are all complex, i.e., $a = a_{i}+ia_{2}$, $b = b_{i}+ib_{2}$, $c = c_{i}+ic_{2}$ and $d = d_{i}+id_{2}$. In this case, all four eigenvalues of the matrix $R$ becomes
\begin{eqnarray}
\lambda_{1}^{'} = \frac{1}{2}\Big[\alpha^{'} + \sqrt{\beta^{'}}\Big]  ~~,~~\lambda_{2}^{'} =\frac{1}{2}\Big[\alpha^{'} - \sqrt{\beta^{'}}\Big] ~~,~~ \lambda_{3}^{'} = (a_{1}^{2}+a_{2}^{2})(d_{1}^{2}+d_{2}^{2})P^{2} ~~,~~\lambda_{4}^{'} = (b_{1}^{2}+b_{2}^{2})(c_{1}^{2}+c_{2}^{2})P^{2}
\end{eqnarray}
Here, the superscript $(')$ indicates that these are the eigenvalues for complex input state parameters case. The expression of $\alpha^{'}$ is
\begin{eqnarray}
\alpha^{'} = |ad-bc|^{2}(2-P)^{2} + 2\Big\{ (a_{1}d_{1}-a_{2}d_{2})(b_{1}c_{1}-b_{2}c_{2})+(a_{1}d_{2}+a_{2}d_{1})(b_{1}c_{2}+b_{2}c_{1})\Big\} P(4-3P)
\end{eqnarray}
where
\begin{eqnarray}
|ad-bc|^{2} = \Big[\Big\{ (a_{1}d_{1}-a_{2}d_{2})-(b_{1}c_{1}-b_{2}c_{2})\Big\}^{2}+\Big\{ (a_{1}d_{2}+a_{2}d_{1})-(b_{1}c_{2}+b_{2}c_{1})\Big\}^{2}\Big]
\end{eqnarray}
and $\beta^{'}$ becomes
   \begin{eqnarray}
   \beta^{'} = \alpha^{'2} - 4\beta^{''}
   \end{eqnarray}
where
\begin{eqnarray}
\beta^{''} = (a_{1}^{2}+a_{2}^{2})(b_{1}^{2}+b_{2}^{2})(c_{1}^{2}+c_{2}^{2})(d_{1}^{2}+d_{2}^{2})P^{2}(4-3P)^{2} = \big|a\big|^{2}\big|b\big|^{2}\big|c\big|^{2}\big|d\big|^{2}P^{2}(4-3P)^{2}
\end{eqnarray}
From the expressions of the eigenvalues, we can easily say that $\lambda_{1}^{'}$ is the largest eigenvalue. At $P=0$, $\beta^{''}=0$ and $\beta^{'} = \alpha^{'2}$. As a result, $\lambda_{1}^{'} =4|ad-bc|^{2}$ and $\lambda_{2}^{'} =\lambda_{3}^{'} =\lambda_{4}^{'} =0$. So, only $\lambda_{1}$ contributes to the concurrence , i.e., $C = \sqrt{\lambda_{1}}=2|(ad-bc)|$. If we increase the channel parameter $P$ gradually, we observe that the eigenvalue $\lambda_{1}$ remains the largest eigenvalue with the increase of $P$ while the other three eigenvalues first become nonzero and then increase gradually. As a result, the concurrence gradually decreases with the increases of $P$. So, in this case, it is clear that we are not getting any rebirth of entanglement or creation of entanglement under the phase damping channel.

    Next, consider the decoherence of an inter-particle entangled state $|\psi\rangle_{in} $ under the phase damping channel. The following three Kraus operators can describe the effect of noise due to phase damping channel \cite{preskill98},
      
      \begin{eqnarray*}
      M_{0}=\sqrt{1-P}\begin{pmatrix}
  1 & 0\\
 0 & 1
\end{pmatrix}
~~,~~
      M_{1}=\sqrt{P}\begin{pmatrix}
  1 & 0\\
 0 & 0
\end{pmatrix}
~~,~~
      M_{2}=\sqrt{P}\begin{pmatrix}
  0 & 0\\
 0 & 1
\end{pmatrix}
      \end{eqnarray*}
  After evolution of $|\psi\rangle_{in}$ through the phase damping channel, the final state becomes
       \begin{eqnarray}
\rho_{out}  = 
\begin{pmatrix}
  a^{2} & ab(1-P) & ac(1-P) & ad(1-P)^{2}\\
  ab(1-P) & b^{2} & bc(1-P)^{2} & bd(1-P)\\
   ac(1-P) & bc(1-P)^{2} & c^{2} & cd(1-P)\\
  ad(1-P)^{2} & bd(1-P) & cd(1-P) & d^{2}
\end{pmatrix}
\end{eqnarray}
To investigate the nature of decoherence of an interparticle entangled state, we consider input state parameters $a=c=\frac{1}{4}$. Fig.~\ref{fig:phaseinter} shows the variation of the concurrence with $P$ for different values of input state parameter $d$. For $d=0$, the concurrence drops to zero when $P=1$. For all non-zero values of $d$, the concurrence becomes zero for the value of $P < 1$.
 
 Fig.~\ref{fig:phasecompare} shows the decoherence of both intraparticle and interparticle entanglement as a function of the channel parameter $P$. Here, the solid and dashed graphs represent the concurrence of the intraparticle entangled state and interparticle entangled state, respectively.  From Fig.~\ref{fig:phasecompare}, one can easily say that interparticle entanglement decays more rapidly than intraparticle entanglement.
 
\begin{figure}[h]
\centering
\includegraphics[scale=0.7]{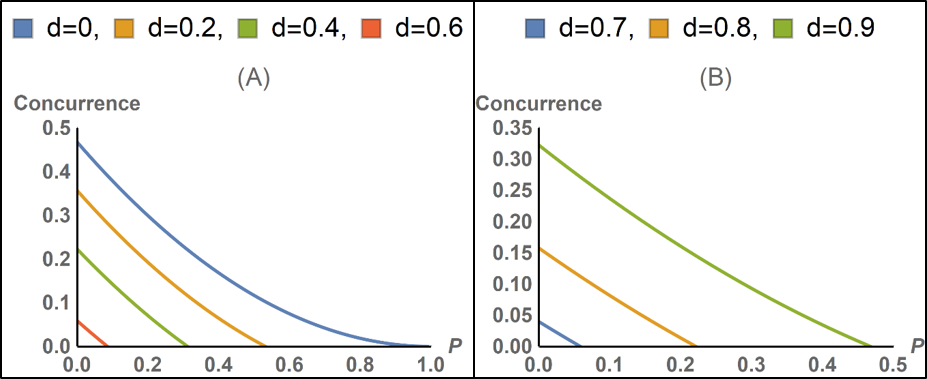}
\caption{Evolution of the concurrence of pure interparticle entangled states with channel parameter $P$ for $a=c=\frac{1}{4}$ and different values of $d$ under the phase damping channel is shown in this figure. In the Fig. 9(A) and Fig. 9(B), $d$ takes the value $0,0.2,0.4,0.6$ and $0.7,0.8,0.9$ respectively. This shows that the concurrence decays non-linearly with the increase of $P$. For $d=0$, the concurrence drops to zero when $P=1$. For all non zero values of $d$, the concurrence becomes zero for value of $P < 1$.}
\label{fig:phaseinter}
\end{figure}

 \begin{figure}[h]
\centering
\includegraphics[scale=0.7]{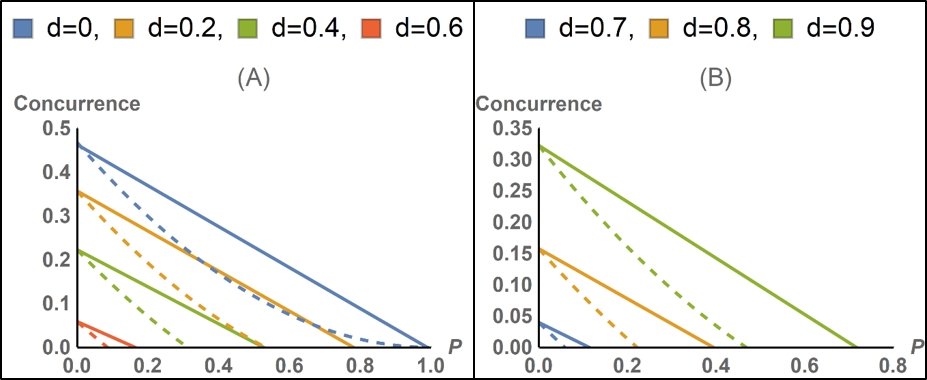}
\caption{Evolution of the concurrence of pure intraparticle entangled states and pure interparticle entangled states with channel parameter $P$ for $a=c=\frac{1}{4}$ and different values of $d$ under the effect of the phase damping channel is shown in this figure. Solid curves represent concurrence of intrparticle entangled states and the dotted curves represent concurrence of interparticle entangled states. In Fig. 10(A) and Fig. 10(B), d takes the value $0,0.2,0.4,0.6$ and $0.7,0.8,0.9$, respectively. Both Fig. 10(A) and Fig. 10(B) show that interparticle entanglement decays more rapidly than intraparticle entanglement under the effect of the phase damping channel.}
\label{fig:phasecompare}
\end{figure}

\section{\label{sec:appendix3}Depolarising channel}
  Under a depolarising channel, a d-level quantum system becomes a maximally mixed state $\mathcal{I}/d$ with finite probability $P$. Kraus operators corresponding to this channel are \cite{AlejandroFonseca, PooladImany, PGokhale}
\begin{eqnarray*}
M_{00}&=&\sqrt{1-\frac{d^{2}-1}{d^{2}}P}~\hat{U}_{00} =\sqrt{1-\frac{15}{16}P}~~\hat{U}_{00},\\
M_{ij}&=&\frac{\sqrt{p}}{4}\hat{U}_{ij} ~,~0\leq  i,j \leq 3 ~~ and ~~ (i,j)\neq (0,0)
\end{eqnarray*}
where $\hat{U}_{ij}$ are the Weyl operators. In a $d$ dimensional Hilbert space, the Weyl operators $\hat{U}_{mn}$ are a set of $d^{2}$ operators defined as \cite{Bertlmann2008,AlejandroFonseca}
     \begin{eqnarray}
     \hat{U}_{mn} = \sum_{j=0}^{d-1}e^{\frac{2\pi i}{d}jm}|j\rangle\langle j+n ~~mod~~ d|
     \end{eqnarray}
Weyl operators are unitary operators and form an orthonormal basis of the Hilbert space, known as the Weyl operator basis. In our present scenario, $d=4$. So
     \begin{eqnarray}
     \hat{U}_{mn} &=& \sum_{j=0}^{3}e^{\frac{2\pi i}{d}jm}|j\rangle\langle j+n ~~mod~~ 4|\nonumber\\
     &=& |0\rangle\langle n|+e^{i\frac{m\pi}{2}}|1\rangle\langle n+1| +e^{im\pi}|2\rangle\langle n+2| \nonumber\\
     &+& e^{i\frac{3m\pi}{2}}|3\rangle\langle n+3|
     \end{eqnarray}
      There are 16 Weyl operators in a $4$ dimensional Hilbert space.

Under the action of a depolarising channel, a d-level quantum system becomes a maximally mixed state $\mathcal{I}/d$ with finite probability $P$. First we consider an input intraparticle entangled state $|\psi\rangle_{in}$ where input state parameters $a$, $b$, $c$ and $d$ are all real. After the evolution of $|\psi\rangle_{in}$ through a depolarising channel, the final state becomes
\begin{eqnarray}
\rho_{out} =\begin{pmatrix}
  a^{2}q+\frac{P}{4} & abq & acq & adq\\
  abq &  b^{2}q+\frac{P}{4} & bcq & bdq\\
   acq & bcq & c^{2}q+\frac{P}{4} & cdq\\
  adq & bdq & cdq & d^{2}q+\frac{P}{4}
\end{pmatrix}
\end{eqnarray}
where $q=1-P$.  Next we calculate the matrix $R$ and find out all four eigenvalues of this matrix. We have provided the definition of $R$ on page number $2$ of the main manuscript. Eigenvalues of $R$ are
\begin{eqnarray}
\label{eqn:concurrence_real1}
\lambda_{1} &=& \alpha + \beta ~,~ \lambda_{2} = \alpha - \beta ~,~ \lambda_{3}=\lambda_{4} = \frac{P^{2}}{16}
\end{eqnarray}
where
\begin{eqnarray}
\label{eqn:concurrence_real2}
\alpha &=& 2(ad-bc)^{2}(1-P)^{2}+\frac{P}{4}(1-\frac{3P}{4}),\nonumber\\
\beta &=& 2|(ad-bc)|(1-P)\sqrt{(ad-bc)^{2}(1-P)^{2}+\frac{P}{4}(1-\frac{3P}{4})}\nonumber\\
\end{eqnarray}
Among four eigenvalues, $ \lambda_{3}$ and $\lambda_{4}$ are input state parameter independent. They depend only on channel parameter $P$. At $P=0$, $\lambda_{3}=\lambda_{4} =0$. With the increase of $P$, they increase and attain their maximum value of $1/16$ at $P=1$. For $a$, $b$, $c$ and $d$ taking positive and negative values, $\alpha$ and $\beta$ are always positive. So, the eigenvalue $\lambda_{1}$ is always greater than $\lambda_{2}$. At $P=0$, no effect of depolarizing noise is there. At that point $\alpha = \beta = 2(ad-bc)^{2}$ and $\lambda_{1}=4(ad-bc)^{2}$, $\lambda_{2}=0$. So, only $\lambda_{1}$ contributes to the expression of the concurrence, i.e., $C = \sqrt{\lambda_{1}}=2|(ad-bc)|$, while the other three eigenvalues have no contribution to the concurrence. If we increase the channel parameter $P$ gradually, we observe that the eigenvalue $\lambda_{1}$ remains the largest eigenvalue with the increase of $P$ while the other three eigenvalues first become nonzero and then increase gradually. At the point when $C=0$, the following condition is satisfied:
\begin{eqnarray*}
\sqrt{\lambda_{1}} = \sqrt{\lambda_{2}}+\sqrt{\lambda_{3}}+\sqrt{\lambda_{4}}
\end{eqnarray*}
Substituting expressions of all $\lambda_{i}$ in the above equation and solving it for $P$ we get following acceptable solution of $P$:
\begin{eqnarray}
\label{eqn:concurrence_real3}
P = \frac{4|(ad-bc)|}{1+4|(ad-bc)|}
\end{eqnarray}
Since $\frac{4|(ad-bc)|}{1+4|(ad-bc)|}<1$, we always get entanglement sudden death in this case. $P$ becomes maximum when one starts from a maximally entangled pure Bell state for which $P=\frac{2}{3}$ when ESD occurs. For nonmaximally entangled pure state $P<\frac{2}{3}$.

      From the above analysis we observe that the eigenvalue $\lambda_{1}$ is always larger than any other three eigenvalues for all possible input state parameters and channel parameter. There is a physical value of the channel parameter $P$ where $\sqrt{\lambda_{1}} = \sqrt{\lambda_{2}}+\sqrt{\lambda_{3}}+\sqrt{\lambda_{4}}$ and we observe ESD. Below the ESD point where $\sqrt{\lambda_{1}} > \sqrt{\lambda_{2}}+\sqrt{\lambda_{3}}+\sqrt{\lambda_{4}}$ we get nonzero Concurrence and above the ESD point where $\sqrt{\lambda_{1}} < \sqrt{\lambda_{2}}+\sqrt{\lambda_{3}}+\sqrt{\lambda_{4}}$ we get zero Concurrence. If we observe the condition $\sqrt{\lambda_{1}} > \sqrt{\lambda_{2}}+\sqrt{\lambda_{3}}+\sqrt{\lambda_{4}}$ after the ESD point, we can say that rebirth of intraparticle entanglement occurs. But, this does not happen for the depolarizing channel.

      In the above analysis, we have considered an intraparticle entangled state whose state parameters $a$, $b$, $c$ and $d$ are all real. Let us now consider an input intraparticle entangled state whose state parameters are all complex, i.e., $a = a_{i}+ia_{2}$, $b = b_{i}+ib_{2}$, $c = c_{i}+ic_{2}$ and $d = d_{i}+id_{2}$. In this case, all four eigenvalues of the matrix $R$ becomes
\begin{eqnarray*}
\label{eqn:concurrence_complex1}
\lambda_{1}^{'} = \alpha^{'} +\beta^{'} ~~,~~ \lambda_{2}^{'} = \alpha^{'} - \beta^{'} ~~,~~ \lambda_{3}^{'} = \lambda_{4}^{'} = \frac{P^{2}}{16}
\end{eqnarray*}
Here the expression of $\alpha^{'}$ and  $\beta^{'}$ become
\begin{eqnarray}
\label{eqn:concurrence_complex2}
\alpha^{'} &=& 2|ad-bc|^{2}(1-P)^{2} +\frac{P}{4}\Big(1-\frac{3P}{4}\Big)\nonumber\\
\beta^{'} &=& 2|ad-bc|(1-P)\sqrt{|ad-bc|^{2}(1-P)^{2}+\frac{P}{4}\Big(1-\frac{3P}{4}\Big)}
\end{eqnarray}
where,
   \begin{eqnarray}
   \label{eqn:concurrence_complex3}
   |ad-bc| = \Big[\Big\{ (a_{1}d_{1}-a_{2}d_{2})-(b_{1}c_{1}-b_{2}c_{2})\Big\}^{2}+\Big\{ (a_{1}d_{2}+a_{2}d_{1})-(b_{1}c_{2}+b_{2}c_{1})\Big\}^{2}\Big]^{1/2}
   \end{eqnarray}
    Here the eigenvalues $\lambda_{3}^{'}$ and $\lambda_{4}^{'}$ are identical to $\lambda_{3}$ and $\lambda_{4}$ because these eigenvalues are input state parameters independent. If we compare  carefully Eq.~(\ref{eqn:concurrence_real2}) and Eq.~(\ref{eqn:concurrence_complex2}), it is clear that, if we replace $|ad-bc|^{2}$ with just a simple square $(ad-bc)^{2}$ we get back from $\alpha^{'}$ to $\alpha$. Similar case for $\beta^{'}$ to $\beta$. Now, at the point where entanglement sudden death occurs,
    \begin{eqnarray*}
\sqrt{\lambda_{1}^{'}} = \sqrt{\lambda_{2}^{'}}+\sqrt{\lambda_{3}^{'}}+\sqrt{\lambda_{4}^{'}}
\end{eqnarray*}
Substituting expressions of all $\lambda_{i}^{'}$ in the above equation and solving it for $P$, we get following physically acceptable solution of $P$:
    \begin{eqnarray}
P = \frac{4|(ad-bc)|}{1+4|(ad-bc)|}
\end{eqnarray}
This value of $P$ looks identical to the value of $P$ in the  Eq.~(\ref{eqn:concurrence_real3}), representing the real input state parameters case. For the real input state parameters case, $|(ad-bc)|$ is the absolute value of the difference of two terms $ad$ and $bc$, while for this case, $|(ad-bc)|$ is given by Eq.~(\ref{eqn:concurrence_complex3}). Since $\frac{4|(ad-bc)|}{1+4|(ad-bc)|}<1$, we always get entanglement sudden death in this case also. As a result, the decay of the concurrence with $P$ will be similar to the real input state parameters case.

   Next, consider an interparticle entangled state under a depolarising channel. In the inter particle entanglement scenario, both particles $A$ and $B$ are spatially separated from each other, and the environment acts on each individual particle separately. So the kraus operators act on the state of each individual particle are two dimensional. They are \cite{Almeida2008}:
\begin{eqnarray}
M_{0}=\sqrt{1-P}\begin{pmatrix}
1 & 0\\
0 & 1
\end{pmatrix}
,
M_{i}=\sqrt{\frac{P}{3}}\sigma_{i} ~~(i=1,2,3)
\end{eqnarray}
where $i=1,2,3$ corresponding to three Pauli operators $\sigma_{x}$, $\sigma_{y}$ and $\sigma_{z}$.
 We consider that the channel parameters for particles $A$ and $B$ are the same and equal to $P$.  After the evolution of $|\psi\rangle_{in}$ through a depolarising channel, the final state concurrence becomes a function of $a$, $c$, $d$ and $P$ where we replace $b$ using the relation $b^{2}=1-a^{2}-c^{2}-d^{2}$. To investigate the decay of interparticle entanglement under the depolarizing channel, consider specific values of $a$ and $c$ as $a=c=\frac{1}{4}$. In this case, the decay of the concurrence with channel parameter $P$ for different values of $D$ is shown in fig.~\ref{fig:depolinter}. From Fig.~\ref{fig:depolinter}, one can say that the concurrence decays linearly with the increase of channel parameter $P$. For all allowed values of $D$, the concurrence becomes zero when $P<1$. So, for all values of $D$, ESD occurs. The higher the initial Concurrence, the longer it takes to decay out under the effect of this noise.
\begin{figure}[h]
\centering
\includegraphics[scale=0.7]{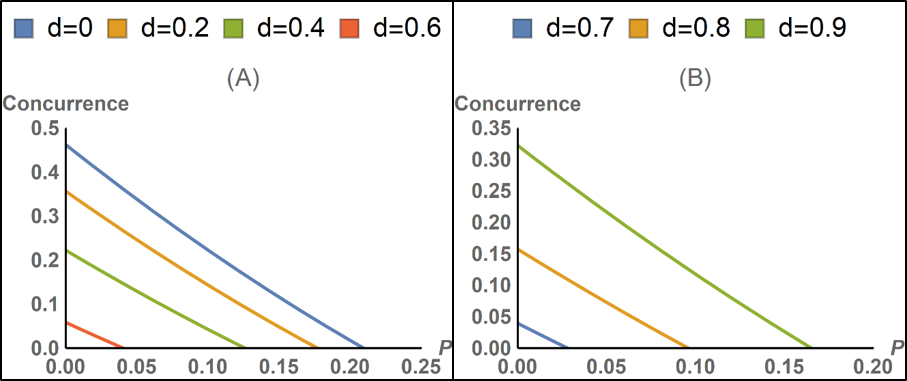}
\caption{Evolution of the concurrence of pure interparticle entangled states with channel parameter $P$ for $a=c=\frac{1}{4}$ and difference values of $d$ under the effect of the depolarising channel is shown in this figure. In Fig. 11(A) and Fig. 11(B), d takes the values $0,0.2,0.4,0.6$ and $0.7,0.8,0.9$, respectively. This figure shows that the concurrence decays linearly with the increase of $P$. For all possible values of $d$, ESD occurs.}
\label{fig:depolinter}
\end{figure}

        Next, we compare the decoherence between an intraparticle entangled state and an interparticle entangled state under a depolarizing channel. So, we consider an intraparticle entangled state and an interparticle entangled state with the same initial Concurrence value and check the nature of the decay of both types of entanglement under the depolarizing channel. Fig.~\ref{fig:depolcompare} shows the decoherence of both types of entanglement. In fig.~\ref{fig:depolcompare}, solid curves represent the concurrence of intraparticle entangled states, while the dashed curves represent the concurrence of interparticle entangled states. From Fig.~\ref{fig:depolcompare}, one can easily say that interparticle entanglement decays more rapidly compared to intraparticle entanglement. So, intraparticle entanglement is more robust than interparticle entanglement under the depolarising channel's effect.
\begin{figure}[h]
\centering
\includegraphics[scale=0.7]{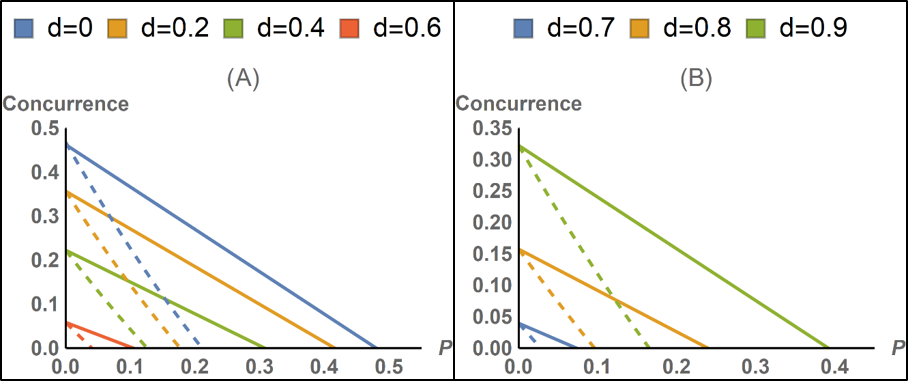}
\caption{Evolution of the concurrence of pure intraparticle entangled states and pure interparticle entangled states with channel parameter $P$ for $a=c=\frac{1}{4}$ and difference values of $d$ under the effect of the depolarising channel is shown in this figure. Solid curves represent concurrence of intrparticle entangled states and the dotted curves represent concurrence of interparticle entangled states. In Fig. 12(A) and Fig. 12(B), d takes the value $0,0.2,0.4,0.6$ and $0.7,0.8,0.9$, respectively. Both the fig. 12(A) and Fig. 12(B) show that interparticle entanglement decays much more rapidly compared to intraparticle entanglement under the effect of the depolarizing channel.}
\label{fig:depolcompare}
\end{figure}

\end{widetext}



\nocite{*}
\begin{thebibliography}{100}%
\bibitem{Horodecki09} R. Horodecki, P. Horodecki, M. Horodecki, and K. Horodecki, \bibinfo{journal}{Rev. Mod. Phys.} \textbf{81}, 865 (2009).
\bibitem{Azzini20}  S. Azzini, S. Mazzucchi, V. Moretti, D. Pastorello, and L.
Pavesi, \bibinfo{journal}{Adv. Quantum Technol.} \textbf{3}, 2000014 (2020).
\bibitem{Basu} S. Basu, S. Bandyopadhyay, G. Kar, D. Home, \bibinfo{journal}{Phys. Lett. A} \textbf{279,} 281 (2001).
\bibitem{Michler2000} M. Michler, H. Weinfurter, M. \.{Z}ukowski, \bibinfo{journal}{Phys. Rev. Lett.} \textbf{84,} 5457 (2000).
\bibitem{Gadway2009} B. R. Gadway, E J Galvez, F. De. Zela, \bibinfo{journal}{J. Phys. B: At. Mol. Opt. Phys.} \textbf{42,} 015503 (2009).
\bibitem{Barreiro2005} J. Barreiro, N. K. Lankford, N. A. Peters, P. G. Kwiat, \bibinfo{journal}{Phys. Rev. Lett.} \textbf{95,} 260501 (2005).
\bibitem{Hasegawa2003} Y. Hasegawa, R. Loidl, G. Badurek, M. Baron, H. Rauch, \bibinfo{journal}{Nature} \textbf{425,} 45 (2003).
\bibitem{Shen2020}  J. Shen, S. J. Kuhn, R. M. Dalgliesh, V. O. de Haan, N. Geerits, A. A. M. Irfan, F. Li, S. Lu, S. R. Parnell, J. Plomp, A. A. van Well, A. Washington, D. V. Baxter, G. Ortiz, W. M. Snow, and R. Pynn, \bibinfo{journal}{Nat. Commun.} \textbf{11,} 930 (2020).
\bibitem{Sun2011} Y. Sun, Q.-Y. Wen, Z. Yuan, \bibinfo{journal}{Opt. Commun.} \textbf{284,} 527 (2011).
\bibitem{Adhikari2015} S. Adhikari, D. Home, A. S. Majumdar, A. K. Pan, Akshata Shenoy H, · R. Srikanth, \bibinfo{journal}{Quantum Inf Process} \textbf{14,} 1451 (2015).
\bibitem{Heo2015} J. Heo, C.-H. Hong, J.-I. Lim, H.-J. Yang, \bibinfo{journal}{Chin. Phys. B} \textbf{24,} 050304 (2015).
\bibitem{Pramanik2010} T. Pramanik, D. Home, S. Adhikari, A. Pan, \bibinfo{journal}{Phys. Lett. A} \textbf{374,} 1121 (2010).
\bibitem{Hong2015} J. Heo, C.-H. Hong, J.-I. Lim, H.-J. Yang, \bibinfo{journal}{Int. J. Theor. Phys.} \textbf{54,} 2261 (2015).
\bibitem{Adhikari2010} S. Adhikari, A. Majumdar, D. Home, A. Pan, \bibinfo{journal}{Eur. Phys. Lett.} \textbf{89,} 10005 (2010).
\bibitem{Yu2009} T. Yu, J. Eberly, \bibinfo{journal}{Science} \textbf{323,} 598 (2009).
\bibitem{Alm2007} M. P. Almeida, E. De MELO, M. Hor-Meyll, A. Salles, S. P. Walborn, P. H. Souto Ribeiro, and L. Davidovich, \bibinfo{journal}{Science} \textbf{316,} 579 (2007).
\bibitem{Almeida2008} A. Salles, F. de Melo, M. P. Almeida, M. Hor-Meyll, S. P. Walborn, P. H. Souto Ribeiro, and L. Davidovich, \bibinfo{journal}{Phys. Rev. A} \textbf{78,} 022322  (2008).
\bibitem{BBellomo} B. Bellomo, R. Lo Franco, and G. Compagno, \bibinfo{journal}{Phys. Rev. Lett.} \textbf{99,} 160502 (2007).
\bibitem{Kraus83} K. Kraus, \textit{States, Effects, and Operations: Fundamental
Notions of Quantum Theory} (Springer-Verlag, Berlin, Heidelberg, 1983).
\bibitem{Wootters1998} William K. Wootters, \bibinfo{journal}{Phys. Rev. Lett.} \textbf{80}, 2245 (1998).
\bibitem{KarlKraus} K. Kraus, States, Effects and Operations: Fundamental Notions of Quantum Theory \bibinfo{journal}{Springer,Berlin,} (1983).


\bibitem{ArijitDutta} A Dutta, J Ryu, W Laskowski, and M Zukowski, \bibinfo{journal}{Phys. Lett. A} \textbf{380,} 2191 (2016).
\bibitem{AlejandroFonseca} A Fonseca, \bibinfo{journal}{Phys. Rev. A} \textbf{100,} 062311 (2019).
\bibitem{Bennett1996} C. H. Bennett, D. P. DiVincenzo, J. A. Smolin, and W. K. Wootters, \bibinfo{journal}{Phys. Rev. A} \textbf{54,} 3824 (1996).
\bibitem{Vidal2000} G. Vidal, \bibinfo{journal}{J. Mod. Opt.} \textbf{47,} 355 (2000).





\bibitem{Brun2002} D. Braun, \bibinfo{journal}{Phys. Rev. Lett. } \textbf{89} 277901  (2002).
\bibitem{Hamdouni2006} Y. Hamdouni, M. Fannes and F. Petruccione, \bibinfo{journal}{Phys. Rev. B } \textbf{73} 245323  (2006).
\bibitem{Ficek2006} Z.Ficek and R. Tanas, \bibinfo{journal}{Phys. Rev. A } \textbf{74} 024304  (2006).
\bibitem{Sun2007} Z. Sun, X. Wang and C. P. Sun, \bibinfo{journal}{Phys. Rev. A } \textbf{75} 062312  (2007).
\bibitem{preskill98} J. Preskill, “Quantum Information and Computation,” California Institute of Technology, 1998 (lecture notes).
 \bibitem{PGokhale} P. Gokhale, J. M. Baker, C. Duckering, N. C. Brown, K. R. Brown, and F. T. Chong, \textit{Proceedings of the 46th International Symposium on Computer Architecture}—ISCA ’19 (ACM, New York, 2019), pp. 554–566.
 \bibitem{PooladImany} P. Imany, J. A. Jaramillo-Villegas, M. S. Alshaykh, J. M. Lukens, O. D. Odele, A. J. Moore, D. E. Leaird, M. Qi, and A. M. Weiner, \bibinfo{journal}{npj Quant. Info.} \textbf{5,}59 (2019).
\bibitem{Bertlmann2008} R. A. Bertlmann and P. Krammer, \bibinfo{journal}{, J. Phys. A} \textbf{41,} 235303 (2008).
\end{thebibliography}
\end{document}